\documentclass[12pt,a4paper]{article}

\pdfoutput=1
\usepackage{jheppub,psfrag,slashed,cancel,lscape,caption,array,graphicx,subcaption}
\usepackage[utf8]{inputenc}
\usepackage{amsmath}
\usepackage{nicefrac}  % for 1/2 
\usepackage{mathtools}
\usepackage{mathrsfs}
\usepackage{multirow}
\usepackage{booktabs} %nice tables 
\usepackage{chngcntr}
\usepackage{nicefrac}
\usepackage{tikz}
\usetikzlibrary{decorations.pathmorphing}
\usetikzlibrary{decorations.markings}
\usetikzlibrary{positioning, shapes, snakes, arrows}

\tikzset{
	graviton/.style={line width=.8pt, -latex,decorate, decoration={snake, segment length=4pt,amplitude=1.8pt, pre length=.1cm, post length=.25cm}},
	worldline/.style={gray, line width=1pt},
	worldlineBold/.style={black, line width=.6pt},
	zUndirected/.style={line width=1pt},
	zParticle/.style={line width=1pt,postaction={decorate},decoration={markings,mark=at position .6 with {\arrow[#1]{latex}}}},
	zParticleF/.style={line width=1pt,postaction={decorate}},
	cscalar/.style={line width=1pt,postaction={decorate},decoration={markings,mark=at position .6 with {\arrow[#1]{latex}}}},
	cscalar2/.style={line width=1pt,postaction={decorate},decoration={markings,mark=at position .8 with {\arrow[#1]{latex}}}},
	photon/.style={line width =.8pt, decorate, decoration={snake, segment length=4pt, amplitude=1.8pt,  pre length=.1cm, post length=.1cm}}
}

\DeclareFontFamily{OT1}{pzc}{}
\DeclareFontShape{OT1}{pzc}{m}{it}{<-> s * [1.350] pzcmi7t}{}
\DeclareMathAlphabet{\mathpzc}{OT1}{pzc}{m}{it}

\def\cN{\mathcal{N}}
\def\cO{\mathcal{O}}

\def\cM{\mathcal{M}}

\def\cS{\mathcal{S}}

\def\eps{\epsilon}

\def\d{\mathrm{d}}

\def\bH{\hat{b}}

\def\dd{\delta\!\!\!{}^-\!}

\def\d{\mathrm{d}}
\def\eps{\epsilon}

\def\tr{\operatorname*{tr}}

\def\braket#1{\langle #1 \rangle}

\def\nn{\nonumber}

\def\eqn#1{eq.~\eqref{#1}}
\def\eqns#1#2{eqs.~\eqref{#1} and~\eqref{#2}}

\def\sec#1{section~{\ref{#1}}}
\def\secs#1#2{sections~{\ref{#1}} and~{\ref{#2}}}

\def\app#1{appendix~{\ref{#1}}}

\def\rcite#1{ref.~\cite{#1}}
\def\rcites#1{refs.~\cite{#1}}

\def\Rcites#1{Refs.~\cite{#1}}

\newcommand{\xdot}{{\dot x}}

\newcommand{\be}{\begin{equation}}
\newcommand{\ee}{\end{equation}}
\newcommand{\ba}{\begin{align}}
\newcommand{\ea}{\end{align}}
\ifx\genfrac\sdflkaj\else\fi
\newcommand{\sfrac}[2]{{\textstyle\frac{#1}{#2}}}

\newcommand{\antic}[2]{#1^{[\mu}#2^{\nu]}}

\newcommand{\ga}{{\mathfrak a}}
\newcommand{\gb}{{\mathfrak b}}
\newcommand{\gc}{{\mathfrak c}}

\begin{document}

\begin{flushright}
\begingroup\footnotesize\ttfamily
	HU-EP-21/28-RTG
\endgroup
\end{flushright}

\vspace{15mm}

\begin{center}
{\LARGE\bfseries 
	SUSY in the Sky with Gravitons
\par}

\vspace{15mm}

\begingroup\scshape\large 
	Gustav Uhre Jakobsen,${}^{1,2}$ Gustav~Mogull,${}^{1,2}$ Jan~Plefka${}^{1}$ and Jan~Steinhoff${}^{2}$  
\endgroup
\vspace{5mm}
					
\textit{${}^{1}$Institut f\"ur Physik und IRIS Adlershof, Humboldt-Universit\"at zu Berlin, \phantom{${}^2$}\\
  Zum Gro{\ss}en Windkanal 2, D-12489 Berlin, Germany} \\[0.25cm]
\textit{${}^{2}$Max-Planck-Institut f\"ur Gravitationsphysik
(Albert-Einstein-Institut)\\
M\"uhlenberg 1, D-14476 Potsdam, Germany } \\[0.25cm]

\bigskip
  
\texttt{\small\{gustav.uhre.jakobsen@physik.hu-berlin.de, gustav.mogull@aei.mpg.de,
jan.plefka@hu-berlin.de, jan.steinhoff@aei.mpg.de\}} 

\vspace{10mm}

\textbf{Abstract}\vspace{5mm}\par
\begin{minipage}{14.7cm}
Picture yourself in the wave zone of a gravitational scattering event of two massive, spinning compact
bodies (black holes, neutron stars or stars).
We show that this system of genuine astrophysical interest enjoys a hidden $\mathcal{N}=2$ supersymmetry, at least to the order of spin-squared (quadrupole) interactions 
in arbitrary $D$ spacetime dimensions.
Using the $\cN=2$ supersymmetric worldline action,
augmented by finite-size corrections for the non-Kerr black hole case,
we build a quadratic-in-spin extension
to the worldline quantum field theory (WQFT) formalism introduced in our previous work,
and calculate the two bodies' deflection and spin kick to sub-leading order
in the post-Minkowskian expansion in Newton's constant $G$.
For spins aligned to the normal vector of the scattering plane we also obtain the scattering angle.
All $D$-dimensional observables are derived from an eikonal phase given as the free energy of the WQFT that is invariant under the $\mathcal{N}=2$  supersymmetry transformations.
\end{minipage}\par

\end{center}
\setcounter{page}{0}
\thispagestyle{empty}
\newpage

\tableofcontents

%**************************************************
\section{Introduction}

Our Universe is populated by massive astrophysical bodies (black holes, neutron stars and stars) that rotate intrinsically --- they carry spin. Along their trajectories through space and time scattering events may
occur that are mediated by the gravitational force. The two scattered bodies will be deflected, their spins will be altered 
and the scattering process will emit gravitational Bremsstrahlung,
which could be detected in future-generation gravitational wave observatories on or near Earth. In this work we show that this
astrophysical system enjoys a hidden, extended ($\mathcal{N}=2$) supersymmetry that constrains the dynamics of the spinning
scattering process --- at least to 
the order of spin-squared (or quadrupole) interactions.
The supersymmetry is manifested by anti-commuting worldline vectors $\psi^{a}$
attributed to the spin tensor of the body
($S^{ab}\sim \bar\psi^{[a}\psi^{b]}$).
These fermionic degrees of freedom are a natural ingredient in the
recently developed worldline quantum field theory (WQFT)
description of massive spinning bodies
\cite{Mogull:2020sak,Jakobsen:2021smu,Jakobsen:2021lvp}.
The appearance of supersymmetry in such a problem of genuine astrophysical interest was first realized in \rcite{Gibbons:1993ap} by studying hidden symmetries of the spinning black hole (Kerr) solution.

The investigation of classical gravitational scattering has a long history in general relativity
~\cite{Kovacs1,Kovacs2,Kovacs:1977uw,Kovacs:1978eu,Bel:1981be,Westpfahl:1985tsl},
being naturally performed in
a post-Minkowskian (PM) perturbative expansion in Newton's constant $G$ about a flat spacetime background
($g_{\mu\nu}=\eta_{\mu\nu}+ \sqrt{32\pi G}\, h_{\mu\nu}$) with the graviton field $h_{\mu\nu}$. 
This \emph{unbound} setup is different to
the intensively studied post-Newtonian (PN) expansion in both $G$ and relative velocity ($\frac{Gm}{r} \sim v^{2}$) often used for binary inspirals,
i.e.~massive bodies on \emph{bound} orbits
with separation $r$. The gravitational radiation emitted in the inspiral finally
leads to a 
merger and is now routinely detected in gravitational wave observatories \cite{Abbott:2016blz,LIGOScientific:2018mvr,LIGOScientific:2020ibl,LIGOScientific:2021usb}. 
Even though Bremsstrahlung-emitting gravitational scattering events appear to be out of reach for current gravitational wave observatories due to their non-periodic signal and lower intensity \cite{Kocsis:2006hq,Mukherjee:2020hnm,Zevin:2018kzq} they represent interesting targets for future searches, calling for precision calculations.

It has also been emphasized recently that gravitational scattering
is relevant for the study of binary inspirals.
There exist various options for mapping between the bound and unbound problems,
including reconstructing a gravitational two-body potential
\cite{Cheung:2018wkq,Neill:2013wsa,Vaidya:2014kza,Damour:2017zjx,Bjerrum-Bohr:2019kec},
or mapping directly between physical observables such as the scattering angle (unbound)
and periastron advance (bound) \cite{Kalin:2019rwq,Kalin:2019inp}.
The scattering problem per se is the natural habitat for quantum field theory (QFT)
that was largely designed to describe the
scattering of elementary particles in collider experiments.
Applying the classical limit of such a perturbative 
Feynman-diagrammatic expansion of the path integral of
matter-coupled gravity in a PM expansion 
has proven to be highly efficient,
and there are two popular QFT-based approaches.

The worldline effective field theory (EFT) formalism
\cite{Goldberger:2004jt,Goldberger:2006bd,Goldberger:2009qd}
models massive bodies as point-like particles moving along their worldlines coupled to the gravitational
field. Spin effects are naturally incorporated by introducing
a spin tensor and co-moving frame along the worldline \cite{Porto:2005ac,Levi:2015msa,Liu:2021zxr}. Similarly, finite-size and tidal effects may be systematically included by coupling the worldline degrees of freedom to higher-dimensional operators 
dressed with Wilson coefficients (or Love numbers) \cite{Kalin:2020lmz}.
Integrating out the graviton fluctuations in the path integral yields an effective action for the interaction of two spinning bodies ---
see \Rcites{Porto:2005ac,Levi:2015msa} for reviews of the PN framework.
Impressively high orders in the PN \cite{Blumlein:2020znm,Bini:2020nsb,Bini:2020hmy,Blumlein:2020pyo,Foffa:2020nqe,Blumlein:2021txj} and PM \cite{Kalin:2020mvi,Kalin:2020fhe,Dlapa:2021npj,Bini:2021gat}
expansions have also been established. 

The second rapidly developing approach is
the amplitudes-based formalism~\cite{Neill:2013wsa,Bjerrum-Bohr:2013bxa,Bjerrum-Bohr:2018xdl,Bern:2019nnu,Bern:2019crd,Cheung:2020gyp,Luna:2017dtq}.
Massive elementary particles (scalars, spin-$\nicefrac{1}{2}$ fermions, etc.)
minimally coupled to the gravitational field are the avatars of black holes,
neutron stars or stars;
their  $2\to 2$ quantum scattering amplitudes are constructed by employing
modern innovations such as generalized unitarity and the double copy, see
\cite{Dixon:1996wi,Elvang:2013cua,Henn:2014yza,Bern:2019prr} for reviews.
Yet, the $2\to 2$ amplitude is only the starting point for a subtle classical limit
\cite{Kosower:2018adc,Maybee:2019jus,Damour:2019lcq}.
This procedure then yields 
the gravitational potential between two massive spinning bodies and observables
such as the deflection and spin kick have been derived from it.
A direct path from amplitudes to classical
observables was introduced in \rcite{Kosower:2018adc}.
Nevertheless, the inclusion of spin poses 
certain challenges, in particular 
going beyond spin-squared interactions \cite{Bern:2020buy} due to the known hurdle of constructing an interacting higher-spin quantum field theory. Tidal and finite-size 
effects have also been included \cite{Bern:2020uwk,Cheung:2020sdj,Aoude:2020ygw}.
Amplitudes-based approaches can employ the powerful tools developed in collider physics and have led to impressively high-order calculations without spin \cite{Bern:2019nnu,Bern:2019crd,Cheung:2020gyp,Bern:2021dqo,Damour:2020tta, DiVecchia:2020ymx,Damour:2019lcq} and first results with spin \cite{Bern:2020buy,Kosmopoulos:2021zoq}
including radiation 
effects~\cite{Amati:1990xe,DiVecchia:2019myk,DiVecchia:2019kta,Bern:2020gjj,AccettulliHuber:2020dal,DiVecchia:2021ndb,Bautista:2019tdr,Herrmann:2021lqe,Herrmann:2021tct}.
A related variant of the amplitude approach is the heavy-particle EFT \cite{Damgaard:2019lfh,Aoude:2020onz,Brandhuber:2021kpo,Brandhuber:2021eyq},
which enables a more straightforward classical limit of the amplitude from the outset.

The WQFT is a new formalism that unites these two approaches and clarifies their interrelations \cite{Mogull:2020sak,Jakobsen:2021smu,Jakobsen:2021lvp}.
The worldline path integral
representation of a massive particle's graviton-dressed propagator corresponds
to the EFT worldline theory's path integral.
Inserting this into the QFT S-matrix (represented as a time-ordered correlation function) a precise connection between the EFT and amplitudes-based approaches
was given for the spinless case \cite{Mogull:2020sak},
which involves the \emph{classical} $\hbar\to0$ limit.
A key feature of the WQFT approach --- distinguishing it from the EFT approach ---
 is that \emph{both} the graviton field 
$h_{\mu\nu}$ \emph{and} the worldline fluctuations about straight-line backgrounds (in the scattering case) are
integrated out in the path integral. In short, the worldline degrees of freedom are also quantized.
This leads to an economic Feynman-diagram-based perturbative solution to the
equations of motion of the matter-graviton system.

In the WQFT, direct access to classical observables, such as the 
spinless particle deflection \cite{Mogull:2020sak} and the time-domain asymptotic gravitational waveform \cite{Jakobsen:2021smu,Jakobsen:2021lvp} emerging from the scattering encounter, has been made. The approach is economic
in the sense that (a) one only computes the classically relevant contributions, circumventing the ``super-classical'' subtleties of the amplitudes-based approach,
and (b) there is no need to determine a non-observable and gauge-dependent effective potential, thereafter solving the resulting equations
of motion perturbatively as is done in the EFT approach.
In fact, the WQFT formalism is to date the only approach used to successfully
construct the asymptotic gravitational waveform of the QFT-based approaches  with or without spin \cite{Jakobsen:2021smu,Jakobsen:2021lvp}.\footnote{See \rcites{Luna:2017dtq,Cristofoli:2021vyo,Mougiakakos:2021ckm} for work on the gravitational Bremsstrahlung \emph{integrand} or the related problem in electrodynamics 
in the amplitudes and EFT based approaches.}

In this paper we report on the inclusion of spin degrees of freedom in the WQFT.
As mentioned above spin is introduced through anti-commuting worldline vectors,
building upon \rcites{Howe:1988ft,Gibbons:1993ap,Bastianelli:2005vk,Bastianelli:2005uy}.
In those works it was shown that the higher-spin $\mathcal{N}/2$ field equations
(in flat space) follow from quantizing an  $\mathcal{N}$-extended supersymmetric
particle augmenting the coordinate vector $x^{\mu}(\tau)$ by $\mathcal{N}$ anti-commuting
vectors $\psi^{\mu}$. In (generic) curved spacetimes this is only possible up to 
$\mathcal{N}=2$ supersymmetry (or spin-1) as we will discuss in \sec{sec:susyWQFT}. This limits our approach to the gravitational scattering of spinning objects
up to interactions of quadratic order in spin (quadrupoles) at present.
In our recent letter \cite{Jakobsen:2021lvp} we have already used this formalism to establish the asymptotic waveform of a spinning gravitational Bremsstrahlung
event, extending the seminal work of Crowley, Kovacs and Thorne \cite{Kovacs1,Kovacs2,Kovacs:1977uw} to the spinning case.

Our spinning WQFT formalism innovates over existing approaches to classical spin based on corotating-frame variables~\cite{Porto:2005ac,Levi:2015msa} in the effective field theory (EFT) of compact objects~\cite{Goldberger:2004jt,Goldberger:2006bd,Goldberger:2009qd,Porto:2016pyg,Levi:2018nxp} ---
see \rcites{Goldberger:2017ogt,Goldberger:2016iau,Shen:2018ebu} for the construction of PM integrands and \rcites{Liu:2021zxr,Kalin:2020mvi} for the computation of
worldline and spin deflections
(shown to be in agreement with the amplitude based results~\cite{Bern:2020buy,Kosmopoulos:2021zoq}).
The EFT approach was applied to radiation also in the PN approximation~\cite{Porto:2010zg,Porto:2012as,Maia:2017gxn,Maia:2017yok,Cho:2021mqw}--- see \rcites{Mishra:2016whh,Buonanno:2012rv} for more traditional methods.
Other approaches to PM spin effects can be found in \rcites{Vines:2017hyw,Bini:2017xzy,Bini:2018ywr,Guevara:2017csg,Vines:2018gqi,Guevara:2018wpp,Chung:2018kqs,Guevara:2019fsj,Chung:2019duq,Damgaard:2019lfh,Aoude:2020onz,Guevara:2020xjx}. 

This paper is organized as follows.
In \sec{sec:susyWQFT} we review the
supersymmetric worldline formalism and give a detailed analysis of the $\mathcal{N}=2$
supersymmetric particle in a curved background. In \sec{sec:routhians} we show
that this theory is equivalent to the traditional description of spinning particles
using spin and body-fixed frame fields,
and explain how finite-size effects may be incorporated
while maintaining the supersymmetry up to the desired quadratic-in-spin order.
In \sec{sec:sec4} we lay out the
spinning WQFT and establish its Feynman rules,
the relationship of the eikonal phase of a scattering event
to the free energy of the spinning WQFT and 
show the  hidden supersymmetry properties of the eikonal phase. 
Finally, in \sec{sec:Examples}
we compute a larger class of observables:
the eikonal phase up to 2PM (next-to-leading) order from which we derive
the deflection and spin kick.
We close with concluding remarks and
in the appendices give details of the $\mathcal{N}=2$ supersymmetry of the gauge-fixed spinning WQFT action and on 
the computation of the necessary integrals arising in \sec{sec:Examples}.

\section{Supersymmetric worldline actions}
\label{sec:susyWQFT}

Extending the WQFT to include spin calls for a worldline
theory of a relativistic spinning particle.
In this section we review the first-order formulation of spinning particle
actions where spin is represented by anti-commuting vector fields.
Our main focus is the $\cN=2$ supersymmetric theory in a generic curved background,
which represents massive spinning bodies up to quadratic order in spin.

\subsection{Putting spin on the worldline}
\label{sec:spinOnWorldline}

To begin with recall the first-order formulation of the massless non-spinning particle in a gravitational background:
\be \label{Spin0flat}
{\tilde S}_{0} = -\int\!\d\tau \left [  p_{M} \dot x^{M} - e \, H\right ]\, , \qquad \text{with}\quad H=\sfrac{1}{2} p_{M}p_{N} g^{MN} \, .
\ee
The transition to the massive case is easily done through a dimensional reduction setting $p^{M}=(p^{\mu},m)$, which yields 
$H=\frac{1}{2}(p^{2}-m^{2})$. We work with mostly minus signature and take the spacetime manifold to be $M^{1,D-1}\times S^{1}$. Eliminating the momentum through its algebraic equations of motion, $p^{\mu}=e^{-1}\dot x^{\mu}$, yields the second-order action
\be\label{Stilde0}
S_{0} = -\int\!\d\tau \left [ \frac{e^{-1}}{2} g_{\mu\nu}\dot x^{\mu}\dot x^{\nu} + \frac{e}{2}m^{2}\right ]\, , 
\ee
which is the ``Polyakov'' formulation of the non-spinning worldline theory enjoying a local reparametrization 
invariance $\tau\to \tau'(\tau)$ under which $e$ transforms as $e\to e \,\dot{\tau}'$. One may also eliminate the einbein $e$ using its algebraic equations of motion to arrive at the usual action for a massive particle $m\int\d s$. However, it is more convenient to choose the proper-time gauge $e=1/m$
as this linearizes the graviton worldline interaction
\cite{Kalin:2020mvi,Mogull:2020sak}.

In order to include spin for the particle one adds real anti-commuting \emph{vector} fields $\psi^{a}_{\alpha}(\tau)$ to the
worldline degrees of freedom \cite{Brink:1976sz,Howe:1988ft,Howe:1989vn}. Here
$\alpha=1,\ldots \mathcal{N}$ is a flavor index while $a=0,\ldots, D-1$ is the flat tangent space index related to the curved 
spacetime index $\mu=0,\ldots, D-1$ via the vielbein $e^{a}_{\mu}(x)$, with $g_{\mu\nu}=e_{\mu}^{a}e_{\nu}^{b}\eta_{ab}$ as usual.
Of particular interest are the supersymmetric worldline theories
which have been shown
to describe the propagation of relativistic quantum fields of spin $\nicefrac{1}{2}$ for $\mathcal{N}=1$ supersymmetry \cite{Brink:1976sz}, spin 1 for $\mathcal{N}=2$ supersymmetry and generally spin $\mathcal{N}/2$ for $\mathcal{N}$-extended supersymmetry \cite{Howe:1988ft,Howe:1989vn} in a \emph{flat}  spacetime background, respectively. 
The $\mathcal{N}$-extended supersymmetric spinning particle generalization of \eqn{Spin0flat} in  a \emph{flat} $(D+1)$-dimensional spacetime
background is\footnote{We take $M=(\mu,D)$ and $A=(a,D)$ with an $S^{1}$ in the $(D+1)$th dimension.}
\begin{align}
\tilde S_{N}= -\int\!\d\tau \left [  p_{M} \dot x^{M} + \sfrac{i}{2} \psi^{A}_{\alpha}\dot\psi^{B}_{\alpha}\eta_{AB}
- e \, H -i\chi^{\alpha} Q_{\alpha} - \frac{1}{2} f_{\alpha\beta} M^{\alpha\beta}\right ]\, 
\end{align}
with the conserved charges
\begin{align}
\label{Ncharges}
H= \sfrac{1}{2} p^{2}\, , \qquad Q_{\alpha}= p\cdot \psi_{\alpha}\, , \qquad M_{\alpha\beta}=i\psi_{\alpha}\cdot \psi_{\beta}\, .
\end{align}
The Lagrange multipliers of the einbein $e(\tau)$, the $\mathcal{N}$ gravitinos $\chi^{a}(\tau)$ and the O($\cN$)  gauge field $f_{\alpha\beta}(\tau)$ enforce
the conservation of these charges. This gauged theory may be thought of as a \emph{locally} supersymmetric worldline theory, i.e.~$\mathcal{N}$-extended 1D supergravity, supersymmetrizing the
reparametrization invariance of the spinless case \eqn{Spin0flat}.
Upon using the Poisson (resp.~Dirac) brackets for the worldline fields
\be
\{ x^{M}, p_{N}\}_{\text{P.B.}} = \delta^{M}_{N}\, , \qquad
\{ \psi^{A}_{\alpha}, \psi^{B}_{\beta}\}_{\text{P.B.}} = -i\delta_{\alpha\beta}\, \eta^{AB}\, ,
\ee
one derives  the  $\mathcal{N}$-extended supersymmetry algebra
\begin{align}
\label{SUSYalgebra}
\{ Q_{\alpha}, Q_{\beta}\}_{\text{P.B.}} &= -2i\delta_{\alpha\beta} H\, , \qquad
\{ H, Q_{\beta}\}_{\text{P.B.}} = \{ H,M_{\alpha\beta}\}_{\text{P.B.}} = 0 \, , \nn \\
\{ M_{\alpha\beta}, Q_{\gamma}\}_{\text{P.B.}} &= -2 \delta_{\gamma[\alpha} Q_{\beta]}\, , \qquad
\{ M_{\alpha\beta}, M^{\gamma\delta}\}_{\text{P.B.}} = -4\delta_{[\alpha}{}^{[\gamma}\, M_{\beta]}{}^{\delta]}\, .
\end{align}
The charges $M_{\alpha\beta}$ generate an O($\cN$) $R$-symmetry algebra.
Again, a dimensional reduction of the theory from $D+1$ to $D$ dimensions, thereby setting $p^{M}=(p^{\mu},m=\text{const})$ and $(\psi^{A}=\psi^{a}_{\alpha},\theta_{\alpha})$, yields the massive theory in $D$ dimensions ---
as in the spinless case of \eqn{Stilde0}. 

The relevant question for our application is whether this theory may be embedded in an \emph{arbitrary curved}  spacetime background whilst preserving supersymmetry.
This is only possible for $\mathcal{N}\leq 2$ describing a spin 0, $\nicefrac{1}{2}$ or spin $1$ particle   in a \emph{generic} curved background \cite{Howe:1988ft,Howe:1989vn,Bastianelli:2002qw,Bastianelli:2005vk,Bastianelli:2005uy}. For the $\mathcal{N}=4$ spinning particle a consistent quantization requires the backgound spacetime to be an Einstein manifold, i.e.~$R_{\mu\nu}=\lambda g_{\mu\nu}$ \cite{Bonezzi:2018box}.
As stated earlier,
our focus is now the spinning $\mathcal{N}=2$ superparticle in a generic curved background,
which allows
us to describe spinning massive bodies up to \emph{quadratic} order in spin.

\subsection{$\mathcal{N}=2$ spinning superparticle in a curved background}
\label{sec:neq2Theory}

Following \rcite{Bastianelli:2005uy} in order to prepare for the massive theory via dimensional reduction we consider the $\mathcal{N}=2$ spinning superparticle
in the curved background $M^{1,D-1}\times S^{1}$. It is convenient to combine the two real Grassmann fields into one complex Grassmann worldline field via $\psi^{A}=\frac{1}{\sqrt{2}}(\psi_{1}^{A}
+i\psi^{A}_{2})$ and $\bar\psi^{A}=\frac{1}{\sqrt{2}}(\psi_{1}^{A}-i\psi^{A}_{2})$. We note the Poisson (resp.~Dirac) brackets
\begin{equation}\label{eq:poissonBrackets}
\{ x^{M}, p_{N}\}_{\text{P.B.}} = \delta^{M}_{N}\, , \qquad\{\psi^{A}, \bar\psi^{B}\}_{\text{P.B.}} = -i\eta^{AB}\, .
\end{equation}
The covariantization of the super-charges \eqref{Ncharges} takes the form
\begin{equation}\label{eq:2charges}
Q= \psi^{a}\, e_{a}^{\mu}(x)\, \pi_{\mu} + \theta\, m\, , \qquad
\bar Q= \bar\psi^{a}\, e_{a}^{\mu}(x)\, \pi_{\mu}+ \bar\theta\, m\, , 
\end{equation}
where we have split off the fifth dimension explicitly $\psi^{5}:=\theta$, $p_{5}=m$ and introduced the covariantized four-momentum
\be
\label{eq:pidef}
\pi_{\mu}:= p_{\mu} -i \omega_{\mu ab} \bar\psi^{a}\psi^{b} \, ,
\ee
with the spin connection $\omega_\mu{}^{ab}=e^a_\nu(\partial_\mu e^{\nu b} + \Gamma^{\nu}_{\mu\lambda}e^{\lambda b})$.
The Hamiltonian may now be derived from the Poisson bracket $\{ Q, \bar Q\}_{\text{P.B.}}$. Using $\{\pi_{\mu}, \pi_{\nu}\}_{\text{P.B.}} =
i R_{\mu\nu ab}\bar\psi^{a}\psi^{b}$ and  $\{\psi^{\mu}, \pi_{\nu}\}_{\text{P.B.}} =
\Gamma^{\mu}_{\nu\rho}\psi^{\rho}$ one finds
\be
\label{Qqbarcurved}
\{ Q, \bar Q\}_{\text{P.B.}} = -2i\Bigl [\underbrace{ \sfrac{1}{2} ( g^{\mu\nu} \pi_{\mu} \pi_{\nu}-m^{2} -
R_{abcd} \bar\psi^{a}\psi^{b} \bar\psi^{c}\psi^{d})}_{H} \Bigr ] -2\underbrace{\pi_{\mu}\bar\psi^{a}\psi^{b}\, T^{\mu}{}_{ab}}_{T}\, .
\ee
Note that the last term $T$ couples to the torsion tensor $T^{\nu}{}_{\mu\rho}:= \Gamma^{\nu}_{[\mu\rho]}$, i.e.~the
antisymmetric part of the affine connection  that vanishes in Einstein gravity. In addition $\{ Q, Q\}_{\text{P.B.}}=0$ vanishes
due to the cyclic identity for the Riemann tensor. Finally, in the complex basis for the Grassmann variables the
internal $R$-symmetry turns into a U(1) symmetry generated by the charge
\be
J=\bar\psi^{a}\psi^{b}\eta_{ab}- \bar\theta\theta \,,
\ee
which obeys $\{ J,  Q\}_{\text{P.B.}}= -iQ$ and $\{ J,  \bar Q\}_{\text{P.B.}}= i\bar Q$.

In summary, the gauged first-order
form of the $\mathcal{N}=2$ superparticle action in a curved spacetime background takes the form
\begin{align}\label{eq:susyAction}
\begin{aligned}
\tilde S&= -\int\!\d\tau \left [ p_{\mu} \dot x^{\mu} + i \bar\psi^{a}\dot\psi^{b}\eta_{ab} - i \bar\theta\dot\theta
- e \, H - i\bar\chi Q - i\chi \bar Q + a\, J\right ] \\
&= -\int\!\d\tau \Bigl [  p_{\mu} \dot x^{\mu} + i \bar\psi^{a}\dot\psi^{b}\eta_{ab} - i \bar\theta\dot\theta
- \frac{e}{2} \left ( 
 g^{\mu\nu} \pi_{\mu} \pi_{\nu}-m^{2}  -
R_{abcd} \bar\psi^{a}\psi^{b} \bar\psi^{c}\psi^{d}
\right ) \\
&  \qquad\qquad
- i \pi_{\mu}( \bar\chi \psi^{\mu} -\bar\psi^{\mu}\chi) - i m(\bar\chi\theta -\bar\theta\chi)
+ a (\bar\psi\cdot \psi -\bar\theta \theta) 
 \Bigr ]\,,
\end{aligned}
\end{align}
with the einbein $e$, the complex gravitino $\chi$ and the U(1) gauge field $a$.
Eliminating $p_{\mu}$ by its algebraic equations of motion yields the action 
\begin{align}
S= -\int\!\d\tau \Bigl [&\frac{e^{-1}}{2}g_{\mu\nu}(\dot x^{\mu}- i\bar\chi \psi^{\mu}- i\chi \bar\psi^{\mu})
(\dot x^{\nu}- i\bar\chi \psi^{\nu}- i\chi \bar\psi^{\nu}) \nn\\
& + i\bar\psi^{a}(\dot\psi_{a}+\dot x^{\mu}\omega_{\mu ab}\psi^{b})  +\frac{e}{2}
R_{abcd}\bar\psi^{a}\psi^{b}\bar\psi^{c}\psi^{d}+\frac{e}{2}m^{2} +a (\bar\psi^{a}\psi_{a}-\bar\theta\theta)\nn\\ & 
- i\bar\theta\dot\theta - im(\bar\chi\theta+\chi\bar\theta)
\Bigr ]\, .
\end{align}
A number of comments are now in order. Firstly, this theory is invariant under \emph{local}
reparametrization, supersymmetry and U(1) gauge symmetries under which both the worldline
fields $x^{\mu}, \psi_{a}$ as well as $e, \chi, \bar\chi, a$ transform ---
see \rcites{Bastianelli:2005vk,Bastianelli:2005uy} for explicit formulae.
Secondly, as shown in \rcite{Bastianelli:2005vk}, one may add a Chern-Simons
term for the gauge field $a$ to the action
\be
S_{CS}=s \int\!\d\tau \, a \, , \quad s=\text{const}\,,
\ee
which is invariant under the U(1) transformation $\delta a =\dot \alpha$, $\delta\psi_{a}=i\alpha\,\psi_{a}$. Finally, one may
set the Lagrange multipliers $\chi$ and $a$ to zero, yielding the constraints
\begin{align}
\label{eq:Constraints}
 \bar \chi=0:& \qquad g_{\mu\nu}\psi^{\mu}\dot x^{\nu}+em\theta=0\, , \quad \qquad \chi=0:\qquad g_{\mu\nu}\bar\psi^{\mu}\dot x^{\nu}+em\bar\theta=0 \, ,\nn\\
a=0:& \qquad \eta_{ab}\bar\psi^{a} \psi^{b} -\bar\theta \theta = s\, .
\end{align}
We shall see in section \ref{sec:routhians} that these may be related to the spin-supplementary condition
(SSC) as well as the conservation of
the total spin in the traditional formulation of describing spinning compact objects.

The resulting second-order theory is then
\begin{align}\label{eq:finalAction}
S^{\cN=2}
= -\!\int\!\d\tau \Bigl [\frac{e^{-1}}{2}g_{\mu\nu}\dot x^{\mu}\dot x^{\nu}
+ i\bar\psi^{a}(\dot\psi_{a}\!+\!\dot x^{\mu}\omega_{\mu ab}\psi^{b})  + \frac{e}{2}
R_{abcd}\bar\psi^{a}\psi^{b}\bar\psi^{c}\psi^{d}+\frac{e}{2}m^{2} %\nn\\ & 
-i\bar\theta\dot\theta 
\Bigr ].
\end{align}
The fermionic extra dimensional contribution $\bar\theta\dot\theta$ is free and couples only to the other fields via the constraints \eqref{eq:Constraints} ---
we drop it from now on.
Moreover, it is convenient to further make the gauge choice $e=1/m$
and rescale the fermions by $\sqrt{m}$: 
 \be
 \psi^{a}\to \sqrt{m}\, \psi^{a} \, , \qquad
 \bar\psi^{a}\to \sqrt{m}\, \bar\psi^{a} \, .
 \ee
 This renders the mass-shell constraint as
 \be
 {\dot x}^{\mu} {\dot x}_{\mu}= 1 + R_{abcd}\bar\psi^{a}\psi^{b}\bar\psi^{c}\psi^{d}\, .
 \ee
 Therefore, strictly speaking $\tau$ is not the proper time in this gauge.
This gauge fixed spinning worldline action \eqref{eq:finalAction} (setting $e=1/m$ and rescaling the fermions)  is invariant under the following
global $\mathcal{N}=2$ supersymmetry transformations:
\begin{align}\label{N=2SUSYcurvedfinal}
\begin{aligned}
\delta x^{\mu}&= ie^{\mu}_{a} (\bar\epsilon\psi^{a} + \epsilon\bar\psi^{a})  \, , 
\quad \\
\delta \psi^{a}& = -\epsilon e^{a}_{\mu} \, \dot x^{\mu} - \delta x^{\mu}\omega_{\mu}{}^{a}{}_{b}\psi^{b}\, ,\quad
\delta \bar\psi^{a}= -\bar\epsilon e^{a}_{\mu} \, \dot x^{\mu} - \delta x^{\mu}\omega_{\mu}{}^{a}{}_{b}\bar\psi^{b}\, ,
\end{aligned}
\end{align}
the analysis of which is relegated to \app{AppA}.
There is also a manifest global U(1) symmetry
\begin{equation}\label{eq:u1}
\delta_{a} \psi^{a}= ia\, \psi^{a}\, , \qquad
\delta_{a} \bar\psi^{a}= -ia\, \bar\psi^{a}\, , \qquad \delta_{a}x^{\mu}=0\, ,
\end{equation}
as well as the remnant of the reparametrization symmetry generated by $H$ of \eqn{Qqbarcurved}, which is given by the commutator of
two supersymmetries \eqref{N=2SUSYcurvedfinal}.

\section{Comparison with a spinning compact body }
\label{sec:routhians}

In this section we demonstrate that the $\cN=2$ spinning superparticle action
\eqref{eq:finalAction} represents an alternative description
of the classical physics of a spinning compact body moving in
a gravitational background, up to and including terms quadratic in the spin (quadrupoles).
For this we need to augment the spinning superparticle action by an additional term
capturing finite-size effects and distinguishing black holes from (neutron) stars.
Surprisingly, supersymmetry is preserved in an approximate sense
under this extension of the theory and may be linked to the spin-supplementary condition (SSC).

\subsection{Traditional worldline action}

The traditional worldline action of a spinning compact body describes 
spin via a body-fixed frame ${\Lambda_A}^\mu(\tau)$ along the worldline
using the flat indices $A, B, \ldots$ that are distinct from the vielbein indices $a,b,\ldots$ from above. The first-order action takes the form \cite{Vines:2016unv,Porto:2016pyg,Levi:2018nxp}
\begin{equation}\label{eq:tetradAction}
 S^{\text{spin}} = - \int\!\d\tau \bigg[ \pi_\mu \dot{x}^\mu + \frac{1}{2} S_{\mu\nu} \underbrace{\Lambda_A{}^\mu  \frac{D \Lambda^{A\nu}}{D\tau}}_{\displaystyle =:\Omega^{\mu\nu}} - \lambda (\pi_\mu \pi^\mu - \mathcal{M}^2) - \chi_\mu S^{\mu\nu} \bigg( \frac{\pi_\nu}{\sqrt{\pi^2}} + \Lambda_{0\nu} \bigg) \bigg] \,.
\end{equation}
Here $\frac{D}{D\tau}:=\dot x^{\mu} \nabla_{\mu}$
is the covariant derivative along the curve
and the ortho-normal Lorentz body-fixed frame ${\Lambda_A}^\mu$ 
satisfies $g_{\mu\nu}{\Lambda_A}^\mu{\Lambda_B}^\nu=\eta_{AB}$.
Using ${\Lambda_A}^\mu$ one may construct the (antisymmetric) angular velocity tensor $\Omega^{\mu\nu}$ as shown above.
The Legendre dual of $\Omega^{\mu\nu}$ is the intrinsic angular momentum or spin tensor $S^{\mu\nu}$.
The Lagrange multipliers $\lambda$ and $\chi_\mu$ enforce the mass-shell
constraints $\pi^2=\cM^2$ and the spin-supplementary condition (SSC) respectively;
the latter arises from the necessity of constraining the antisymmetric $S^{\mu\nu}$
to carry only the physical rotational degrees of freedom of the compact body (i.e.~three angles in 4d). If we gauge fix the time-like component of the body-fixed frame ${\Lambda_0}^{\mu}=\pi^\mu/\sqrt{\pi^2}$ then
the SSC takes the covariant form $S_{\mu\nu}\pi^\nu=0$.
Instead, we find it more convenient to gauge fix via the choice $\chi_\mu=0$ for the Lagrange multiplier (analogous to fixing $e$ above) so we may drop the last term in the action, which approximately is equivalent to the covariant SSC $S_{\mu\nu}\pi^\nu=0 + \mathcal{O}(S^3)$.

Starting at quadratic order in spins and linear order in curvature, the parity-invariant mass-shell constraint $\pi^2=\cM^2$
receives curvature couplings  \cite{Vines:2016unv}
\begin{align}
\label{eq:massshellconstr}
	\mathcal{M}^2
	&= m^2 - \sfrac14 R_{\mu\nu\alpha\beta} S^{\mu\nu} S^{\alpha\beta} +
  C_{E} E_{\mu\nu} S^{\mu\rho} P_{\rho\sigma} S^{\nu\sigma}
	+\cO(S^3)\,,
\end{align}
where $C_E$ is a Wilson coefficient
induced by finite-size deformations (for a Kerr black hole $C_E=0$).\footnote{In the literature one often finds $C_{E}\to C_{E}-1$.}
We have introduced the ``electric'' curvature tensor $E_{\mu\nu}$ and projector $P_{\mu\nu}$ as 
\begin{align}
	E_{\mu\nu} &= R_{\mu\alpha\nu\beta} \pi^\alpha \pi^\beta / m^2 \,, \label{Edef} \\
  P_{\mu\nu} &= g_{\mu\nu} - \pi_\mu \pi_\nu / \pi^2 \,. \label{Pdef}
\end{align}
In four spacetime dimensions one may introduce the
Pauli-Lubanski spin vector
 $
 S^\mu = \frac{1}{2} \epsilon^{\mu\nu\alpha\beta} S_{\alpha\beta} \pi_\nu / m 
$
and the additional term takes the form
$
 C_{E} E_{\mu\nu} S^{\mu} S^\nu
$.
Finally, $S^{\mu\nu}S_{\mu\nu}=2s^2$ is a constant along the worldline.

\subsection{Identification with the $\mathcal{N}=2$ spinning particle}

Identification of the traditional action for a spinning compact body \eqn{eq:tetradAction}
with the $\cN=2$ supersymmetric worldline theory of \eqn{eq:finalAction} is
made by identifying the spin tensor using
\begin{equation}\label{eq:spinTensor}
	\boxed{S^{\mu\nu}=-2ie_a^\mu e_b^\nu\bar{\psi}^{[a}\psi^{b]}\,.}
\end{equation}
We have $S^{ab}=e^a_\mu e^b_\nu S^{\mu\nu}$ in a local Minkowski frame. So in a sense
the worldline fermions $\psi^{a}$ are the quarks of the spin field $S^{ab}$.
This identification is well-motivated by noting that the SO(1,3) Lorentz algebra:
\begin{equation}
  \{S^{ab},S^{cd}\}_{\rm P.B.}=
  \eta^{ac}S^{bd}+\eta^{bd}S^{ac}-\eta^{bc}S^{ad}-\eta^{ad}S^{bc}\,,
\end{equation}
follows from $\{\psi^a,\bar\psi^b\}_{\text{P.B.}}=-i\eta^{ab}$ and provides the normalization in \eqn{eq:spinTensor}.
Note that the mass-shell constraint \eqn{eq:massshellconstr} for $C_E=0$ directly maps
with \eqref{eq:spinTensor} to
the Hamiltonian of the $\cN=2$ theory \eqref{eq:susyAction}, i.e.~the Kerr black hole.
Upon (seemingly) sacrificing supersymmetry the finite-size $C_E$ term may
also be included by adding
\be
\label{eq:SESSdef}
H_{E}:= m\,C_EE_{ab}\,  \bar\psi^a \psi^b\, P_{cd}\bar\psi^{c}\psi^{d}
\ee
to the spinning superparticle Hamiltonian $H$ in \eqn{eq:susyAction}.
Using \eqn{eq:spinTensor} one can easily show that this agrees
with the corresponding  term in
the spinning particle action \eqref{eq:tetradAction}.
 We will discuss the
implications of this for SUSY at the end of this section.

How do we prove the equivalence of our $\cN=2$ worldline SUSY theory \eqref{eq:susyAction},
augmented by \eqn{eq:SESSdef},
with the traditional formulation of \eqn{eq:tetradAction}?
Taking inspiration from \rcites{Porto:2016pyg,Liu:2021zxr},
we compare the \emph{Routhians} ${\cal R}$ of the two theories.
The Routhian to be considered is a Legendre transform of the Lagrangian
with respect only to the spin degrees of freedom:
it is a Hamiltonian with respect to the spin,
but a Lagrangian with respect to the position of the particle.
In the traditional spin formalism \eqref{eq:tetradAction} switching to the Routhian
conveniently eliminates all dependence on angular velocity tensor $\Omega^{\mu\nu}$:
\begin{align}\label{eq:spinRouthian}
\begin{aligned}
	{\cal R}^{\rm spin}:\!&=-\frac12S_{ab}\,\Lambda_A^a\dot{\Lambda}^{A\,b}-L^{\rm spin}\\
	&=\pi_\mu \dot{x}^\mu-\frac12\omega_{\mu,ab}\dot{x}^\mu S^{ab} + 
	\lambda (\pi_\mu \pi^\mu - \mathcal{M}^2)\,,
\end{aligned}
\end{align}
where we used that $\Lambda^{a}_{A}\frac{D\Lambda^{A\, b}}{D\tau}=\Lambda^{a}_{A}\, \dot{\Lambda}^{A\, b} - \omega_{\mu}{}^{ab}\dot{x}^\mu$.
Comparing this to the spinning superparticle action in \eqn{eq:susyAction} augmented by 
the finite size term \eqref{eq:SESSdef} the corresponding $\cN=2$ SUSY Routhian takes the form
\begin{align}
\begin{aligned}
  {\cal R}^{\cN=2}:\!&=-i\eta_{ab}\bar{\psi}^a\dot{\psi}^b-L^{\cN=2} 
  - L^{E}= p_{\mu}\dot x^{\mu}
  + e\, (H +  H_{E})\\
  &= \pi_{\mu} \dot x^{\mu} 
  +i\omega_{\mu,ab}\dot{x}^\mu\bar{\psi}^a\psi^b \\ &\qquad
  -\frac{e}{2} \left ( 
   \pi_{\mu} \pi^{\mu}-m^{2}  +
  R_{abcd} \bar\psi^{a}\psi^{b} \bar\psi^{c}\psi^{d}
  +m\,C_EE_{ab}\,  \bar\psi^a \psi^b\, P_{cd} \, \bar\psi^c \psi^d
  \right )\,,
\end{aligned}
\end{align}
having set $\theta=\bar{\theta}=0$ and used the definition of $\pi^{\mu}$ of \eqn{eq:pidef} in the last step.
Upon identifying $\lambda=\sfrac{e}2$ we have a perfect match
${\cal R}^{\rm spin}={\cal R}^{\cN=2}$, thereby also reproducing the finite-size
term coupling to $C_{E}$ in $\cM$ of ${\cal R}^{\rm spin}$.
We also note that the evolution of a generic function $F(\psi^a,\bar{\psi}^a)$
of the spinors is given by Hamilton's equation:
\begin{equation}\label{eq:hamiltonEq}
\frac{\d F}{\d\tau}=\{F,{\cal R}\}_{\rm P.B.}\,.
\end{equation}
Choosing $F$ as $S^{\mu\nu}$ we see that ${\cal R}^{\rm spin}={\cal R}^{\cN=2}$
guarantees consistency of the spin tensor's equation of motion between these two
descriptions.

From our perspective,
the Routhian has now served its purpose and we will not need it again.
Our approach is therefore qualitatively different from other
EFT-based methods, e.g.~\rcite{Liu:2021zxr},
wherein one proceeds by solving
Hamilton's equation \eqref{eq:hamiltonEq} for $S^{\mu\nu}$.
In that context, as we have discussed,
the main benefit of the Routhian \eqref{eq:spinRouthian}
over the Lagrangian \eqref{eq:tetradAction} is that it
does not depend on the angular velocity tensor $\Omega^{\mu\nu}$.
However, by introducing the Grassmann vector $\psi^a$
at an early stage we have already gained this benefit.
We will therefore continue to profit from our use of a fully Lagrangian-based formalism,
using the SUSY action \eqref{eq:susyAction} augmented by the SUSY-breaking term 
\eqref{eq:SESSdef},
treating position and spin of the particles on an equal footing.
This will allow us to perform calculations using a set of Feynman rules,
to be derived in \sec{sec:sec4}.

\subsection{Approximate supersymmetry of the finite-size term}
\label{sect:approxsusy}

Finally, we discuss the implications of adding the SUSY-breaking
finite-size term $H_{E}$ \eqref{eq:SESSdef} to the Hamiltonian.
In fact, this term \emph{does preserve SUSY approximately}
in the sense that the SUSY variation of it vanishes up to terms irrelevant
for spin-squared interactions. This may be seen from the bracket of $H_{E}$ with the (undeformed)
supercharge $Q$: as $H_{E}$ is quartic in fermions the leading  term
in the bracket is cubic in fermions and reads
\be
\{ Q , H_{E}\}_{\text{P.B.}} = mC_{E} (
\underbrace{\,E_{\mu\nu}\pi^{\mu}}_{=0}\, e^{\nu}_{b} \psi^{b} P_{cd}\bar\psi^{c}\psi^{d}
+ E_{ab}\psi^{a}\, \psi^{b} \underbrace{\, \pi^{\gamma}P_{\gamma\rho}}_{=0} e^{\rho}_{d}\psi^{d})
+\cO(\psi^{5})\,,
\ee
which vanishes by virtue of eqs.~\eqref{Edef} and \eqref{Pdef}. Hence the full Hamiltonian
$H+H_{E}$ is \emph{approximately} supersymmetric. The SUSY-violating terms 
$\cO(\psi^{5})$ above will in turn receive corrections from  the SUSY variation
of putative spin-cubed additions to the Hamiltonian.

In fact, one may also incorporate $H_{E}$ into the SUSY algebra \eqref{SUSYalgebra}
in this approximate sense.
Deforming the SUSY generators
$Q$ and $\bar Q$ by $\cO(\psi^{5})$ terms via 
\be
Q\to  Q'= Q+ \frac{\psi\cdot \pi}{\pi^{2}} H_{E}\, , \qquad 
\bar Q\to \bar Q'= \bar Q+ \frac{\bar\psi\cdot \pi}{\pi^{2}}H_{E}\, , \qquad 
\ee
gives rise to the Poisson brackets
\begin{align}
\{ Q' , \bar Q'\}_{\text{P.B.}} & =  -2i H +   
\{ Q ,\frac{\bar\psi\cdot \pi}{\pi^{2}}H_{E} \}_{\text{P.B.}} +
\{ \bar Q ,\frac{\psi\cdot \pi}{\pi^{2}}H_{E} \}_{\text{P.B.}} + \cO(\psi^{6}) \nn \\
&= -2i (H+ H_{E}) + \cO(\psi^{6})\, ,
\end{align}
and $\{ Q' , Q'\}_{\text{PB}}=\cO(\psi^{6})$. Hence, we have an approximate closure of the $\mathcal{N}=2$ SUSY algebra.
As our analysis is valid only up to spin-squared order we may neglect these higher-order
fermionic terms and violations. 
This entails that the SUSY constraints --- alias SSC ---
hold only approximately for $C_E \neq 0$,
i.e.~we have $\pi\cdot\psi=\pi\cdot\bar\psi=0 + \mathcal{O}(\psi^5)$ and $S_{\mu\nu}\pi^\nu=0 + \mathcal{O}(S^3)$.
In \app{AppA} we discuss the Lagrangian version of this result in the gauge-fixed theory.

Finally, the U(1) $R$-symmetry is preserved by the finite-size extension. Having set the Lagrange multiplier $a$ to zero
we also have the constraint $\bar\psi\cdot\psi=s$, which corresponds
the conservation of spin length $S_{\mu\nu}S^{\mu\nu}=2s^2$.

\section{Spinning supersymmetric WQFT}
\label{sec:sec4}

In this section we use the $\cN=2$ SUSY worldline action of \sec{sec:neq2Theory}
to build a spinning generalization of the WQFT formalism \cite{Mogull:2020sak},
valid up to quadratic order in the spins.
For each massive body $i$ we start from the worldline action \eqref{eq:finalAction}:
\begin{align}\label{eq:action}
	S^{(i)}= -m_i\int\!\d\tau \left[\frac12g_{\mu\nu}\dot x_i^{\mu}\dot x_i^{\nu}
	+i\bar\psi_{i,a}\frac{D\psi_i^a}{D\tau}+\frac12
	R_{abcd}\bar\psi_i^{a}\psi_i^{b}\bar\psi_i^{c}\psi_i^{d}\right]\,,
\end{align}
where $\frac{D\psi_i^a}{D\tau}=\dot\psi_i^a+\dot x^\mu{\omega_\mu}^{ab}\psi_{i,b}$
and include the finite-size corrections \eqref{eq:SESSdef}
\begin{equation}\label{eq:finiteSizeDef}
  S^{(i)}_{E}:=
  -m_i\, C_{E,i}\int\!\d\tau\,  R_{a\mu b\nu}\dot{x}_i^\mu\dot{x}_i^\nu \bar\psi_i^a \psi_i^b\, P_{cd}\,\bar\psi_i^{c}\psi_i^{d}\,,
\end{equation}
with two distinct Wilson coefficients $C_{E,1}$ and $C_{E,2}$.
Note that the projector reads $P_{ab}=\eta_{ab}-e_{a \mu}e_{b \nu}\xdot^{\mu}\xdot^{\nu}/\xdot^{2}$ here.
The two bodies interact via ordinary general relativity,
appropriately described by the $D$-dimensional Einstein-Hilbert action and gauge-fixing term:
\begin{align}\label{eq:SEH}
  S_{\rm EH}=-\frac2{\kappa^2}\int\!\d^Dx\sqrt{-g}\,R\,, &&
  S_{\rm gf}=
  \int\!\d^Dx\big(\partial_\nu h^{\mu\nu}-\sfrac12\partial^\mu{h^\nu}_\nu\big)^2\,,
\end{align}
where $\kappa=\sqrt{32\pi G}$ with Newton's constant $G$;
the harmonic gauge-fixing term enforces
$\partial_\nu h^{\mu\nu}=\sfrac12\partial^\mu{h^\nu}_\nu$.

In order to describe a scattering encounter we
expand the worldline fields about their undeflected straight-line trajectories:
\begin{align}
  x_i^\mu(\tau) = b_i^\mu + v_i^\mu \tau + z_i^\mu(\tau)\,, &&
  \psi^a_i(\tau) = \Psi^a_i+{\psi'}_i^a(\tau)\,,
\end{align}
with the new fields $z_i^\mu(\tau)$ and ${\psi'}_i^a(\tau)$ as perturbations.
We also introduce the constant part of the spin tensor $S_i^{\mu\nu}(\tau)$
in the local frame:
\begin{equation}
  {\cal S}_i^{ab}=-2i\bar\Psi_i^{[a}\Psi_i^{b]}\,.
\end{equation}
The bulk metric is expanded around a flat Minkowski vacuum:
\begin{align}
  g_{\mu\nu}(x) &= \eta_{\mu\nu}+\kappa h_{\mu\nu}(x)\,,
\end{align}
using a mostly minus metric signature $\eta_{\mu\nu}={\rm diag}(+1,-1,-1,-1)$.
The corresponding expansions of the vielbein $e_\mu^a$
and spin connection $\omega_\mu{}^{ab}$ are
\begin{subequations}
\begin{align}
  e^{a}_{\mu} &
  = \eta^{a\nu}\left(\eta_{\mu\nu}+ \sfrac{\kappa}{2}h_{\mu\nu} - 
  \sfrac{\kappa^2}{8}h_{\mu\rho}{h^\rho}_\nu  + \cO(\kappa^3) \right)\, ,
  \\
  \omega_\mu{}^{ab} 
  &= -\kappa \partial^{[a} h^{b]}{}_\mu -
  \sfrac{\kappa^{2}}{2} h^{\nu [a}(\partial^{b]}h_{\mu\nu} -\partial_{\nu}h^{b]}{}_{\mu}
    +\sfrac{1}{2}\partial_{\mu}h^{b]}{}_{\nu}) + \mathcal{O}(\kappa^3)\, .
\end{align}
\end{subequations}
In this perturbative framework we no longer distinguish between
curved $\mu,\nu,\ldots$ and tangent $a,b,\ldots$ indices.
The background parameters $b_i^\mu$, $v_i^\mu$ and $\cS_i^{\mu\nu}$,
along with the masses of the two bodies $m_i$,
constitute the physical data regarding the scattering in question. 
We may set $b\cdot v_i=0$, where $b^\mu=b_2^\mu-b_1^\mu$ is the relative impact parameter,
and $v_i\cdot\Psi_i=v_i\cdot\bar\Psi_i=0$, implying $\cS^{\mu\nu}_iv_{i\mu}=0$,
without loss of generality ---
more on this in \sec{sect:sym}.

In the WQFT framework $h_{\mu\nu}(x)$, $z_i^\mu(\tau_i)$
and ${\psi'}_i^\mu(\tau_i)$ are promoted to quantum fields.
The quantum theory is defined by the partition function:
\begin{align}\label{ZWQFTdef}
\begin{aligned}
  \mathcal{Z}_{\text{WQFT}}=e^{i\chi}
  := \text{const} &\times
  \int D[h_{\mu\nu}]
    \int \prod_{i=1}^2 D [z_{i},\psi^\prime_i,\ga_{i},\gb_{i},\gc_{i}]\\
    &\times\exp\Bigl[i\Big(S_{\rm EH}+S_{\rm gf}+
    \sum_{i=1}^2\big(S^{(i)}+S_{E}^{(i)}+S_{\rm ghost}^{(i)}\big)\Big)\Bigr ]\,,
\end{aligned}
\end{align}
where $\chi$ is the \emph{eikonal phase};
the overall constant ensures that ${\cal Z}_{\rm WQFT}=1$
in the non-interacting case ($\kappa=0$), so $\chi=0$.
The additional terms $S^{(i)}_{\rm ghost}$ arise
from the need to write down a covariant path integral measure \cite{Bastianelli:1992ct}:
\begin{align}
\begin{aligned}
\mathcal D [x] &= D[x] \prod_{0\leq\sigma\leq T}\sqrt{-\text{det} g_{\mu\nu}[x(\tau)] } \\ & 
= D[x]
\int D[\ga,\gb,\gc]\, \exp\Bigl[\underbrace{ -i\int_{0}^T\!\d \tau\, \left (\sfrac{1}{2} g_{\mu\nu} 
(\ga^{\mu}\ga^{\nu} + \gb^{\mu}\gc^{\nu}) \right )}_{iS_{\rm ghost}} \Bigr ]
\, ,
\end{aligned}
\end{align}
which brings the ``Lee-Yang'' ghosts $\ga_i^{\mu}$ (commuting) and $\gb_i^{\mu}$, $\gc_i^{\mu}$ (anti-commuting) into the theory.
On top, dimensional regularization of the path integral induces a finite counter term in terms of the Ricci scalar evaluated along the worldline trajectory $-\frac{1}{2}R[x(\tau)]$ \cite{Bastianelli:1992ct,Bastianelli:2000pt}.  However, both of these additions turn out to be irrelevant for the classical considerations that we are interested in in this work.
Observables are calculated as expectation values in the WQFT:
\begin{align}\label{eq:expValues}
\begin{aligned}
    \Bigl \langle \cO(h,\{x_i,\psi_i\})\Bigr \rangle
    :=\mathcal{Z}_{\text{WQFT}}^{-1}
    \int D[h_{\mu\nu}]
    \int \prod_{i=1}^2 D [z_{i},\psi^\prime_i,\ga_{i},\gb_{i},\gc_{i}] \,
    \cO(h,\{x_i,\psi_i\})&\\ 
    \qquad\times\exp\Bigl[i\Big(S_{\rm EH}+S_{\rm gf}+
    \sum_{i=1}^2\big(S^{(i)}+S_{E}^{(i)}+S_{\rm ghost}^{(i)}\big)\Big)\Bigr ]\,,&
\end{aligned}
\end{align}
To straightforwardly compute both these observables and the eikonal phase $\chi$
in practice we derive a set of momentum-space Feynman rules.
These build on the non-spinning rules already derived in \rcite{Mogull:2020sak},
and have already been  presented in \rcite{Jakobsen:2021lvp}.

\subsection{Feynman rules}
\label{sec:feynmanRules}

As the Feynman rules originating from the Einstein-Hilbert action \eqref{eq:SEH}
are fairly conventional we do not dwell on them here;
the graviton is simply re-expressed in momentum space as
\begin{align}
  h_{\mu\nu}(x)=\int_k e^{ik\cdot x}h_{\mu\nu}(-k)\,,
\end{align}
where $\int_k:=\int\!\sfrac{\d^Dk}{(2\pi)^D}$
(the negative sign convention gives Feynman vertices with outgoing momenta).
Spatial integration $\int\!\d^Dx$ in \eqn{eq:SEH} gives rise to
momentum-conserving delta functions
at each $n$-point momentum-space graviton vertex;
also including the gauge-fixing term $S_{\rm gf}$
we read off the graviton propagator from the two-point function:
\begin{align}
  \begin{tikzpicture}[baseline={(current bounding box.center)}]
    \coordinate (x) at (-.7,0);
      \coordinate (y) at (0.5,0);
    \draw [photon] (x) -- (y) node [midway, below] {$k$};
    \draw [fill] (x) circle (.08) node [above] {$\mu\nu$};
    \draw [fill] (y) circle (.08) node [above] {$\rho\sigma$};
  \end{tikzpicture}=i\frac{P_{\mu\nu;\rho\sigma}}{k^2+i\eps}\,, &&
  P_{\mu\nu;\rho\sigma}:=
  \eta_{\mu(\rho}\eta_{\sigma)\nu}-\sfrac1{D-2}\eta_{\mu\nu}\eta_{\rho\sigma}\,.
\end{align}
The Feynman $i\eps$ prescription used here is consistent
with purely conservative scattering,
and will be sufficient at the 2PM order we shall be working at.

Next we consider the supersymmetric worldline action \eqref{eq:action}.
The quadratic terms in $z_i^\mu$ and ${\psi_{i}^{\prime}}^\mu$ are
\begin{align}
	\left.S^{(i)}\right|_{\rm quadratic}&=
  -m_i\int\!\d\tau\,\eta_{\mu\nu}\!\left[\frac12\dot z_i^\mu\dot z_i^{\nu}
	+i\bar{\psi}_i^{\prime\mu}\dot\psi_i^{\prime\nu}\right]\,.
\end{align}
Writing the worldline fields in energy space,
\begin{align}\label{eq:1dfourier}
  z_i^\mu(\tau)=\int_\omega e^{i\omega\tau}z_i^\mu(-\omega)\,, &&
  {\psi_{i}^{\prime}}^\mu(\tau)=
  \int_\omega e^{i\omega\tau}{\psi_{i}^{\prime}}^\mu(-\omega)\,,
\end{align}
where $\int_\omega:=\int\sfrac{\d\omega}{2\pi}$,
we read off the two worldline propagators:
\begin{align}\label{eq:wlPropagators}
  \begin{tikzpicture}[baseline={(current bounding box.center)}]
    \coordinate (in) at (-0.6,0);
    \coordinate (out) at (1.4,0);
    \coordinate (x) at (-.2,0);
    \coordinate (y) at (1.0,0);
    \draw [zUndirected] (x) -- (y) node [midway, below] {$\omega$};
    \draw [dotted] (in) -- (x);
    \draw [dotted] (y) -- (out);
    \draw [fill] (x) circle (.08) node [above] {$\mu$};
    \draw [fill] (y) circle (.08) node [above] {$\nu$};
  \end{tikzpicture}&=-i\frac{\eta^{\mu\nu}}{2m_i}
    \left(\frac1{(\omega+i\eps)^2}+\frac1{(\omega-i\eps)^2}\right)\,,\\
  \begin{tikzpicture}[baseline={(current bounding box.center)}]
    \coordinate (in) at (-0.6,0);
    \coordinate (out) at (1.4,0);
    \coordinate (x) at (-.2,0);
    \coordinate (y) at (1.0,0);
    \draw [zParticle] (x) -- (y) node [midway, below] {$\omega$};
    \draw [dotted] (in) -- (x);
    \draw [dotted] (y) -- (out);
    \draw [fill] (x) circle (.08) node [above] {$\mu$};
    \draw [fill] (y) circle (.08) node [above] {$\nu$};
  \end{tikzpicture}&=-i\frac{\eta^{\mu\nu}}{2m_i}
    \left(\frac1{\omega+i\eps}+\frac1{\omega-i\eps}\right)\,.
\end{align}
As explained in \rcite{Mogull:2020sak}, 
the choice of $i\eps$ prescription determines the
precise interpretation of the background parameters:
with advanced or retarded prescriptions
$b_{i,\pm\infty}^\mu$, $v_{i,\pm\infty}^\mu$
and $\Psi_{i,\pm\infty}^\mu$ are identified with
the undeflected particle trajectories in the limit $\tau\to\pm\infty$.
To leading order in $G$ the time-symmetric $i\eps$ prescription
used here averages over these two possibilities,
hence $\Psi_i^\mu=\sfrac12(\Psi_{i,+\infty}^\mu+\Psi_{i,-\infty}^\mu)+\cO(G^2)$,
$v_i^\mu=\sfrac12(v_{i,+\infty}^\mu+v_{i,-\infty}^\mu)+\cO(G^2)$
and $b_i^\mu=\sfrac12(b_{i,+\infty}^\mu+b_{i,-\infty}^\mu)+\cO(G^2)$.

Finally, we read off interaction vertices from the
supersymmetric worldline action \eqref{eq:action}
and finite-size term \eqref{eq:finiteSizeDef}.
The worldline fields are written in energy space \eqref{eq:1dfourier};
as the graviton coupling to the worldline implicitly depends on $z_i^\mu$ we re-write it thus:
\begin{align}
  h_{\mu\nu}(x_i(\tau))&=
  \int_ke^{ik\cdot(b_i+v_i\tau+z_i(\tau))}h_{\mu\nu}(-k)
  =\sum_{n=0}^\infty\frac{i^n}{n!}\!\int_ke^{ik\cdot(b_i+v_i\tau)}
  \left(k\cdot z_i(\tau)\right)^nh_{\mu\nu}(-k)\nn\\
  &=\sum_{n=0}^\infty\frac{i^n}{n!}\!\int_{k,\omega_1,\ldots,\omega_n}
  e^{ik\cdot b_i}e^{i(k\cdot v_i+\sum_{j=1}^n\omega_j)\tau}
  \left(\prod_{j=1}^nk\cdot z_i(-\omega_j)\right)h_{\mu\nu}(-k)\,.
\end{align}
Integration of the actions over proper time
gives rise to energy-conserving delta functions $\dd(\omega):=2\pi\delta(\omega)$,
and we read off Feynman vertices from the interaction terms.
The simplest vertex is the single-graviton source
(suppressing the $i$ subscripts):
\begin{align}\label{eq:vertexH}
  \begin{tikzpicture}[baseline={(current bounding box.center)}]
    \coordinate (in) at (-1,0);
    \coordinate (out) at (1,0);
    \coordinate (x) at (0,0);
    \node (k) at (0,-1.3) {$h_{\mu\nu}(k)$};
    \draw [dotted] (in) -- (x);
    \draw [dotted] (x) -- (out);
    \draw [graviton] (x) -- (k);
    \draw [fill] (x) circle (.08);
  \end{tikzpicture}&=
  -i\frac{m\kappa }{2}e^{ik\cdot b}\dd(k\cdot v)\\[-10pt]
  &\quad\times\bigg(
  v^\mu v^\nu
  +i (k\cdot\cS)^{(\mu}v^{\nu)}
  -\frac12 (k\cdot\cS)^{\mu}(k\cdot\cS)^{\nu}
  +\frac{C_E}{2}v^\mu v^\nu(k\cdot\cS\cdot\cS\cdot k)\bigg)\,,\nn
\end{align}
where $(k\cdot\cS)^\mu:=k_\nu\cS^{\nu\mu}$,
representing the linearized (in $h_{\mu\nu}$) stress-energy tensor.
In the non-spinning case ($\cS^{\mu\nu}=0$) this precisely reproduces
the corresponding vertex of \rcite{Mogull:2020sak};
for a Kerr black hole ($C_E=0$) the expression is consistent with
an exponential representation of the linearized 
stress-energy tensor \cite{Vines:2017hyw,Guevara:2018wpp,Guevara:2019fsj}
apparently valid to all orders in spin:
\begin{align}
  &h_{\mu\nu}(-k)T^{\mu\nu}(k)=
  me^{ik\cdot b}\dd(k^2)\dd(k\cdot v)(v\cdot\eps)^2
  \exp\left(i\frac{k\cdot\cS\cdot\eps}{v\cdot\epsilon}\right)\\
  &\qquad=me^{ik\cdot b}\dd(k^2)\dd(k\cdot v)\eps_\mu\eps_\nu\!
  \left(v^\mu v^\nu+i(k\cdot\cS)^{(\mu}v^{\nu)}-
  \frac12 (k\cdot\cS)^{\mu}(k\cdot\cS)^{\nu}\right)+\cO(\cS^3)\,,\nn
\end{align}
where the on-shell graviton is $h_{\mu\nu}(k)=\dd(k^2)\eps_\mu\eps_\nu$.

As the higher-point vertices become rapidly more complicated we provide
only the ones required to compute the 2PM eikonal phase in \sec{sec:eikonal}.
Firstly, the graviton coupling to a single deflection mode is
\begin{align}\label{eq:vertexHZ}
  &\begin{tikzpicture}[baseline={(current bounding box.center)}]
     \coordinate (in) at (-1,0);
     \coordinate (out) at (1,0);
     \coordinate (x) at (0,0);
     \node (k) at (0,-1.3) {$h_{\mu\nu}(k)$};
     \draw (out) node [right] {$z^\rho(\omega)$};
     \draw [dotted] (in) -- (x);
     \draw [zUndirected] (x) -- (out);
     \draw [graviton] (x) -- (k);
     \draw [fill] (x) circle (.08);
   \end{tikzpicture}=
  \frac{m\kappa }{2}e^{ik\cdot b}\dd(k\cdot v+\omega)\\
  &\times
  \bigg(2\omega v^{(\mu}\delta^{\nu)}_\rho+v^\mu v^\nu k_\rho
  +i(k\cdot\cS)^{(\mu}(k_\rho v^{\nu)}+\omega\delta_\rho^{\nu)})+
  \frac12k_\rho(k\cdot\cS)^{\mu}(\cS\cdot k)^{\nu}\nn\\
  &\,\,+\frac{C_E}{2}\Big(\left(2\omega v^{(\mu}\delta^{\nu)}_\rho+v^\mu v^\nu k_\rho\right)(k\cdot\cS\cdot\cS\cdot k)
  -\omega^2k_\rho(\cS\cdot\cS)^{\mu\nu}+
  2\omega^2(k\cdot\cS\cdot \cS)^{(\mu}\delta^{\nu)}_\rho\Big)\bigg)\,,\nn
\end{align}
which again reproduces the non-spinning case when $\cS^{\mu\nu}=0$.
The coupling to a single Grassmann-odd vector is
\begin{align}\label{eq:vertexHS}
    &\begin{tikzpicture}[baseline={(current bounding box.center)}]
      \coordinate (in) at (-1,0);
      \coordinate (out) at (1,0);
      \coordinate (x) at (0,0);
      \node (k) at (0,-1.3) {$h_{\mu\nu}(k)$};
      \draw (out) node [right] {$\psi^{\prime\rho}(\omega)$};
      \draw [dotted] (in) -- (x);
      \draw [zParticle] (x) -- (out);
      \draw [graviton] (x) -- (k);
      \draw [fill] (x) circle (.08);
     \end{tikzpicture}\!\!\!\!\!=
    -im\kappa e^{ik\cdot b}\dd(k\cdot v+\omega)\\[-15pt]
    &\qquad\qquad\times
    \bigg(k_{[\rho}\delta_{\sigma]}^{(\mu}\!\left(v^{\nu)}-i(\cS\cdot k)^{\nu)}\right)+
    iC_E\!\left(v^{(\mu}k_\lambda+\omega\delta^{(\mu}_\lambda\right)\!
    \left(v^{\nu)}k_{[\rho}+\omega\delta^{\nu)}_{[\rho}\right)\!{\cS^\lambda}_{\sigma]}
      \bigg)\bar{\Psi}^\sigma\,.\nn
\end{align}
The vertex with an outgoing ${\bar\psi}^{\prime\rho}(\omega)$ line is the same,
except with $\bar{\Psi}^\sigma\to\Psi^\sigma$.
Finally, we require the two-graviton emission vertex from the worldline:
\begin{align}\label{eq:vertexHH}
  &\begin{tikzpicture}[baseline={(current bounding box.center)}]
    \coordinate (in) at (-1,0);
    \coordinate (out) at (1,0);
    \coordinate (x) at (0,0);
    \node (k1) at (-.9,-1.3) {$h_{\mu_1\nu_1}(k_1)$};
    \node (k2) at (.9,-1.3) {$h_{\mu_2\nu_2}(k_2)$};
    \draw [dotted] (in) -- (x);
    \draw [dotted] (x) -- (out);
    \draw [graviton] (x) -- (k1);
    \draw [graviton] (x) -- (k2);
    \draw [fill] (x) circle (.08);
   \end{tikzpicture}=
  -\frac{m\kappa^2}{4}e^{i(k_1+k_2)\cdot b}\dd((k_1+k_2)\cdot v)\\
  &\times
  \bigg(
  (k_1\cdot\cS)^{\mu_2}v^{\mu_1}\eta^{\nu_1\nu_2}-
  \cS^{\mu_1\mu_2}\!\left(v^{\nu_1}k_1^{\nu_2}-
  \sfrac12k_1\cdot v\eta^{\nu_1\nu_2}\right)
  \nn\\  &\quad
  +i\big(
  (\cS\cdot k_1)^{\mu_1}(\cS\cdot k_1)^{\mu_2}
  +\sfrac12(\cS\cdot k_2)^{\mu_1}(\cS\cdot k_1)^{\mu_2}
  -\sfrac12\cS^{\mu_1\mu_2}(k_1\cdot\cS\cdot k_2)
  \big)
  \eta^{\nu_1\nu_2}
  \nn\\  &\quad
  +\sfrac{i}4k_1\cdot k_2\cS^{\mu_1\nu_2}\cS^{\mu_2\nu_1}
  -ik_1^{\nu_2}(\cS\cdot(k_1+k_2))^{\mu_1}\cS^{\mu_2\nu_1}
  \nn\\  &\quad
  +i\,C_E
  \Big(
  \big(
  2k_1\cdot v(\cS\cdot\cS\cdot(k_1+k_2))^{\mu_2}v^{\mu_1}
  -\sfrac12(k_1\cdot v)^2(\cS\cdot\cS)^{\mu_1\mu_2}
  \nn\\  &\quad
  -\sfrac12(k_1\cdot\cS\cdot\cS\cdot k_2)v^{\mu_1}v^{\mu_2}
  \big)
  \eta^{\nu_1\nu_2}
  -\sfrac12k_1\cdot k_2(\cS\cdot\cS)^{\nu_1\nu_2}v^{\mu_1}v^{\mu_2}
  \nn\\  &\quad
  +k_1^{\nu_2}(\cS\cdot\cS\cdot k_2)^{\nu_1}v^{\mu_1}v^{\mu_2}
  -k_1^{\nu_2}(\cS\cdot\cS\cdot k_1)^{\mu_2}v^{\mu_1}v^{\nu_1}
  -k_1^{\nu_2}(\cS\cdot\cS\cdot k_2)^{\mu_2}v^{\mu_1}v^{\nu_1}
  \nn\\  &\quad
  -(\cS\cdot\cS)^{\mu_2\nu_2}\left(k_1\cdot vk_2^{\nu_1}-\sfrac12k_1\cdot k_2v^{\nu_1}\right)v^{\mu_1}
  \Big)
  \bigg)+(1\leftrightarrow2)\,,\nn
\end{align}
which is implicitly symmetrized on $(\mu_1,\nu_1)$ and $(\mu_2,\nu_2)$.

\subsection{Recursive properties}
\label{sec:recursive}

The Feynman rules \eqref{eq:vertexH}, \eqref{eq:vertexHZ} and \eqref{eq:vertexHS}
satisfy recursive properties:
\begin{subequations}
\begin{align}
  \begin{tikzpicture}[baseline={(current bounding box.center)}]
    \coordinate (in) at (-1,0);
    \coordinate (out) at (1,0);
    \coordinate (x) at (0,0);
    \node (k) at (0,-1.3) {$h_{\mu\nu}(k)$};
    \draw (out) node [right] {$z^\rho(0)$};
    \draw [dotted] (in) -- (x);
    \draw [zUndirected] (x) -- (out);
    \draw [graviton] (x) -- (k);
    \draw [fill] (x) circle (.08);
  \end{tikzpicture}&=
  \frac{\partial}{\partial b^\rho}\quad
  \begin{tikzpicture}[baseline={(current bounding box.center)}]
    \coordinate (in) at (-1,0);
    \coordinate (out) at (1,0);
    \coordinate (x) at (0,0);
    \node (k) at (0,-1.3) {$h_{\mu\nu}(k)$};
    \draw [dotted] (in) -- (x);
    \draw [dotted] (x) -- (out);
    \draw [graviton] (x) -- (k);
    \draw [fill] (x) circle (.08);
  \end{tikzpicture}\,,\label{eq:recursiveA}\\
  \begin{tikzpicture}[baseline={(current bounding box.center)}]
    \coordinate (in) at (-1,0);
    \coordinate (out) at (1,0);
    \coordinate (x) at (0,0);
    \node (k) at (0,-1.3) {$h_{\mu\nu}(k)$};
    \draw (out) node [right] {$\psi^{\prime\rho}(0)$};
    \draw [dotted] (in) -- (x);
    \draw [zParticle] (x) -- (out);
    \draw [graviton] (x) -- (k);
    \draw [fill] (x) circle (.08);
  \end{tikzpicture}&=
  \frac{\partial}{\partial\Psi^\rho}\quad
  \begin{tikzpicture}[baseline={(current bounding box.center)}]
    \coordinate (in) at (-1,0);
    \coordinate (out) at (1,0);
    \coordinate (x) at (0,0);
    \node (k) at (0,-1.3) {$h_{\mu\nu}(k)$};
    \draw [dotted] (in) -- (x);
    \draw [dotted] (x) -- (out);
    \draw [graviton] (x) -- (k);
    \draw [fill] (x) circle (.08);
  \end{tikzpicture}\,.\label{eq:recursiveB}
\end{align}
\end{subequations}
In \rcite{Mogull:2020sak} (the non-spinning case)
the first relationship was generalized to $n$ points:
\begin{align}
  \begin{tikzpicture}[baseline={(current bounding box.center)}]
  \coordinate (in) at (-1,0);
  \coordinate (out1) at (1,0);
  \coordinate (out2) at (1,0.5);
  \coordinate (out3) at (1,0.9);
  \coordinate (out4) at (1,1.4);
  \coordinate (x) at (0,0);
  \node (k) at (0,-1.3) {$h_{\mu\nu}(k)$};
  \draw (out1) node [right] {$z^{\rho_1}(\omega_1)$};
  \draw (out2) node [right] {$\!\!\!\vdots$};
  \draw (out3) node [right] {$z^{\rho_n}(\omega_n)$};
  \draw (out4) node [right] {$z^{\rho_{n+1}}(0)$};
  \draw [dotted] (in) -- (x);
  \draw [zUndirected] (x) -- (out1);
  \draw [zUndirected] (x) to[out=30,in=180] (out3);
  \draw [zUndirected] (x) to[out=60,in=180] (out4);
  \draw [graviton] (x) -- (k);
  \draw [fill] (x) circle (.08);
  \end{tikzpicture}=
  \frac{\partial}{\partial b^{\rho_{n+1}}}\quad
  \begin{tikzpicture}[baseline={(current bounding box.center)}]
    \coordinate (in) at (-1,0);
    \coordinate (out1) at (1,0);
    \coordinate (out2) at (1,0.5);
    \coordinate (out3) at (1,0.9);
    \coordinate (x) at (0,0);
    \node (k) at (0,-1.3) {$h_{\mu\nu}(k)$};
    \draw (out1) node [right] {$z^{\rho_1}(\omega_1)$};
    \draw (out2) node [right] {$\!\!\!\vdots$};
    \draw (out3) node [right] {$z^{\rho_n}(\omega_n)$};
    \draw [dotted] (in) -- (x);
    \draw [zUndirected] (x) -- (out1);
    \draw [zUndirected] (x) to[out=30,in=180] (out3);
    \draw [graviton] (x) -- (k);
    \draw [fill] (x) circle (.08);
    \end{tikzpicture}\,.
\end{align}
In words: a vertex with $(n+1)$ external $z^\mu$ particles,
and $\omega_{n+1}=0$,
is given by a derivative with respect to the impact parameter
$b^\mu$ of the corresponding $n$-point vertex.
We claim this continues to hold when spin is included,
and that \eqn{eq:recursiveB} generalizes similarly,
regardless of what other external lines are present on the vertex.
In the non-spinning case we confirmed this recursive property
using an analytic expression for the worldline vertices:
\begin{align}
  &\begin{tikzpicture}[baseline={(current bounding box.center)}]
    \coordinate (in) at (-1,0);
    \coordinate (out1) at (1,0);
    \coordinate (out2) at (1,0.5);
    \coordinate (out3) at (1,0.9);
    \coordinate (x) at (0,0);
    \node (k) at (0,-1.3) {$h_{\mu\nu}(k)$};
    \draw (out1) node [right] {$z^{\rho_1}(\omega_1)$};
    \draw (out2) node [right] {$\!\!\!\vdots$};
    \draw (out3) node [right] {$z^{\rho_n}(\omega_n)$};
    \draw [dotted] (in) -- (x);
    \draw [zUndirected] (x) -- (out1);
    \draw [zUndirected] (x) to[out=30,in=180] (out3);
    \draw [graviton] (x) -- (k);
    \draw [fill] (x) circle (.08);
    \end{tikzpicture}=
  i^{n-1}m\kappa\,
  e^{ik\cdot b}\dd\bigg(k\cdot v+\sum_{i=1}^n\omega_i\bigg)\times\\[-20pt]
  &\qquad\qquad\,\,
  \left(\frac12\left(\prod_{i=1}^nk_{\rho_i}\!\right)\!v^\mu v^\nu+
  \sum_{i=1}^n\omega_i\left(\prod_{j\neq i}^nk_{\rho_j}\right)\!
  v^{(\mu}\delta^{\nu)}_{\rho_i}+
  \sum_{i<j}^n\omega_i\omega_j\left(\prod_{l\neq i,j}^nk_{\rho_l}\right)
  \delta^{(\mu}_{\rho_i}\delta^{\nu)}_{\rho_j}\right)\!.\nn
  \end{align}
With the inclusion of spin, however,
we no longer have such a compact expression and therefore argue differently.

At the Lagrangian level these properties follow straightforwardly from
\begin{align}\label{eq:lRecursive}
  \frac{\partial L^{(i)}(\tau)}{\partial b_i^\mu}=
  \frac{\partial L^{(i)}(\tau)}{\partial z_i^\mu(\tau)}\,, &&
  \frac{\partial L^{(i)}(\tau)}{\partial \Psi_i^\mu}=
  \frac{\partial L^{(i)}(\tau)}{\partial \psi_i^{\prime\mu}(\tau)}\,,
\end{align}
where $S^{(i)}+S_{E}^{(i)}=\int\!\d\tau\,L^{(i)}(\tau)$
(we are now ignoring the ghosts).
The former is true simply because the Lagrangian $L^{(i)}$ depends on $b_i^\mu$ and $z_i^\mu$ only
implicitly via $x_i^\mu(\tau)=b_i^\mu+\tau v_i^\mu+z_i^\mu(\tau)$;
the latter because $L^{(i)}$ depends on spin only via
$\psi_i^\mu(\tau)=\Psi_i^\mu+\psi_i^{\prime\mu}(\tau)$
(and its complex conjugate).
In energy space the action therefore generically depends on
$x_i^\mu(\omega_j)=\dd(\omega_j)b_i^\mu-i\dd^{\,\prime}(\omega_j)v_i^\mu+z_i^\mu(\omega_j)$
and $\psi_i^\mu(\omega_j)=\dd(\omega_j)\Psi_i^\mu+\psi_i^{\prime\mu}(\omega_j)$
for the collection of energies $\{\omega_j\}$.
So, in this case a derivative with respect to the background parameter
$b_i^\mu$ or $\Psi_i^\mu$ is equivalent to one with
respect to the corresponding perturbation
$z_i^\mu(\omega_j)$ or $\psi_i^{\prime\mu}(\omega_j)$,
if we set $\omega_j=0$ (as implied by the delta function).
In the next section we use these properties to
obtain observables from the eikonal phase $\chi$.

\subsection{Observables from the eikonal phase}
\label{sec:eikonal}

The eikonal phase $\chi$ is a scalar with a privileged role in the WQFT,
containing knowledge of both the classical impulse $\braket{\Delta p_{i,\mu}}$
and spin kick $\braket{\Delta S_{i,\mu\nu}}$.
To recover these observables we use the recursive properties of the
worldline vertices \eqref{eq:lRecursive};
from the action
$S=S_{\rm EH}+S_{\rm gf}+\sum_{i=1}^{2}\int\d\tau\,L^{(i)}(\tau)$
these may be re-expressed as
\begin{subequations}\label{eq:eL}
\begin{align}
  \frac{\partial S}{\partial b_i^\mu}&=
  \int_{-\infty}^\infty\!\d\tau
  \left(\frac{\partial L^{(i)}}{\partial x_i^\mu(\tau)}-
  \frac{\d}{\d\tau}\frac{\partial L^{(i)}}{\partial \dot{x}_i^\mu(\tau)}\right)-
  \underbrace{\big[p_{i,\mu}\big]_{\tau=-\infty}^{\tau=\infty}}_{\Delta p_{i,\mu}}\,,\\
  \frac{\partial S}{\partial \Psi_i^\mu}&=
  \int_{-\infty}^\infty\!\d\tau
  \left(\frac{\partial L^{(i)}}{\partial \psi_i^\mu(\tau)}-
  \frac{\d}{\d\tau}\frac{\partial L^{(i)}}{\partial \dot{\psi}_i^\mu(\tau)}\right)+
  im_i\underbrace{\big[\bar\psi_{i,\mu}\big]_{\tau=-\infty}^{\tau=\infty}}_{\Delta\bar\psi_{i,\mu}}\,,
\end{align}
\end{subequations}
where $p_{i,\mu}=-\frac{\partial L^{(i)}}{\partial\dot{x}_i^\mu}$ and
$im_i\bar{\psi}_{i,\mu}=\frac{\partial L^{(i)}}{\partial\dot{\psi}_i^\mu}$.
As the first terms define the classical (Euler-Lagrange)
equations of motion for $x_i^\mu$ and $\psi_i^{\mu}$,
 their expectation values in the WQFT vanish. 
Therefore, by taking derivatives of
the free energy $-i\log\mathcal{Z}_{\text{WQFT}}=\chi$ of \eqn{ZWQFTdef}  
with respect to $b_i^\mu$, $\Psi_i^\mu$ and $\bar\Psi_i^\mu$
and using \eqn{eq:eL} we see that\footnote{Fermionic derivatives act to the right.}
\begin{subequations}
\begin{align}
	\braket{\Delta p_{i,\mu}}&=-\frac{\partial\chi}{\partial b_i^\mu}\,,\\
  im_i\braket{\Delta\psi_{i,\mu}}&=
	\frac{\partial\chi}{\partial\bar\Psi_i^\mu}=
	-2i\Psi_i^\nu\frac{\partial\chi}{\partial{\cal S}_i^{\mu\nu}}\,,\\
	im_i\braket{\Delta\bar\psi_{i,\mu}}&=
	\frac{\partial\chi}{\partial\Psi_i^\mu}=
	-2i\bar\Psi_i^\nu\frac{\partial\chi}{\partial{\cal S}_i^{\mu\nu}}\,.
\end{align}
\end{subequations}
We have used the fact that
$\chi$ depends on $\Psi_i^\mu$ and $\bar\Psi_i^\mu$ only implicitly via
${\cal S}_i^{\mu\nu}=-2i\bar\Psi_i^{[\mu}\Psi_i^{\nu]}$.
The expectation value of the spin kick $\Delta S_i^{\mu\nu}$
is therefore recovered as
\begin{align}\label{eq:spinKickderiv}
\begin{aligned}
	\braket{\Delta S_i^{\mu\nu}}
	&=
	-2i\bar\Psi_i^{[\mu}\braket{\Delta\psi_i^{\nu]}}
	-2i\braket{\Delta\bar\psi_i^{[\mu}}\Psi_i^{\nu]}\\
	&=
	\frac4{m_i}{\cal S}_i^{\rho[\mu}
	\frac{\partial\chi}{\partial{{\cal S}_{i,\nu]}}^\rho}\,.
\end{aligned}
\end{align}
For any field perturbed around a background expectation value,
the expected ``kick'' of that field is therefore extracted from the eikonal phase
by taking a derivative with respect to the corresponding background parameter.

In the special case of aligned spins to the scattering plane
we can also determine the scattering angle $\theta$,
given in terms of the momentum impulse as
\begin{align}\label{eq:angleCoM}
  \sin\left(\frac\theta2\right)=\frac{|\Delta p_i|}{2p_\infty}\,, &&
  p_{\infty}=\frac{m_1m_2}E\sqrt{\gamma^2-1}\,,
\end{align}
where $|\Delta p_i|:=\sqrt{-\braket{\Delta p_i}^2}$,
$p_\infty$ is the center-of-mass momentum,
the total energy is $E=\sqrt{m_1^2+m_2^2+2\gamma m_1m_2}$ and $\gamma = v_{1}\cdot v_{2}$.
From the eikonal phase the scattering angle is directly recovered as
\begin{equation}
  \sin\!\left(\frac\theta2\right)=
  -\frac1{2p_\infty}\frac{\partial\chi}{\partial|b|}\,.
\end{equation}
Using these relations, in \secs{sec:1PMobservables}{sec:2PMobservables} we will calculate
the momentum impulse, spin kick and aligned-spin scattering angle
at 1PM and 2PM order respectively.

\subsection{Background field symmetries}\label{sect:sym}

Invariance of the action under the SUSY transformations \eqref{N=2SUSYcurvedfinal} (see appendix \ref{AppA} for details),
the U(1) symmetry \eqref{eq:u1} and translation invariance along the worldline
has physical consequences for these observables derived from the eikonal phase.
After integrating out the worldline fluctuations $z_i^\mu$ and $\psi_i'^\mu$,
for each transformation there is a flat-space
remnant of these symmetries on the background parameters:
\begin{subequations}\label{eq:shiftSymmetries}
\begin{align}
\delta b_i^\mu&=\xi_i v_i^\mu +i\bar\eps_i\Psi_i^\mu+i\eps_i\bar\Psi_i^\mu\,,\\
\delta v_i^\mu&=0\,,\\
\delta\Psi_i^\mu&=-\eps_i v_i^\mu-i\alpha_i\Psi_i^\mu\,,\\
\delta\bar\Psi_i^\mu&=-\bar\eps_i v_i^\mu+i\alpha_i\bar\Psi_i^\mu\,,
\end{align}
\end{subequations}
for constant shift parameters $\xi_i$, $\eps_i$, $\bar\eps_i$ and $\alpha_i$.
Hence the eikonal phase $\chi$ depending only on the background parameters
will be invariant under
\begin{align}
\delta\chi&=\frac{\partial\chi}{\partial b_i^\mu}\delta b_i^\mu
+\frac{\partial\chi}{\partial\Psi_i^\mu}\delta\Psi_i^\mu
+\frac{\partial\chi}{\partial\bar\Psi_i^\mu}\delta\bar\Psi_i^\mu =0
\end{align}
for both $i=1,2$.
For each parameter we recover a constraint:
\begin{subequations}
\begin{align}
  \xi_i: \quad 0&=v_i^\mu\braket{\Delta p_{i,\mu}}\,,\label{eq:conA}\\
  \eps_i: \quad 0&=\braket{\Delta p_{i,\mu}}\bar\Psi_i^\mu
    +m_iv_i^\mu\braket{\Delta\bar\psi_{i,\mu}}\,,\label{eq:conB}\\
  \bar\eps_i: \quad 0&=\braket{\Delta p_{i,\mu}}\Psi_i^\mu
    +m_iv_i^\mu\braket{\Delta\psi_{i,\mu}}\,,\label{eq:conC}\\
  \alpha_i: \quad 0&=\bar\Psi_i^\mu\braket{\Delta\psi_{i,\mu}}
    +\braket{\Delta\bar\psi_{i,\mu}}\Psi_i^\mu\,.\label{eq:conD}
\end{align}
\end{subequations}
These four constraints respectively imply conservation of
$p_i^2$, $p_i\cdot\bar\psi_i$, $p_i\cdot\psi_i$ and $\psi_i\cdot\bar\psi_i$
between initial and final asymptotic states,
i.e.~the energy/mass, conserved supercharges and spin length.\footnote{
  Although the true supercharges $Q_i=\pi_i\cdot\psi_i$,
  $\bar Q_i=\pi_i\cdot\bar\psi_i$ involve the covariantized momentum
  $\pi_{i,\mu} = p_{i,\mu} -i \omega_{\mu ab} \bar\psi_i^{a}\psi_i^{b}$,
  the spacetime is asymptotically flat so $\omega_{\mu ab}=0$ at the boundary.
}

How do we re-interpret the latter three constraints in terms of the
classical spin tensors $S_i^{\mu\nu}$?
Using $\braket{\Delta S_i^{\mu\nu}}=
-2i\bar\Psi_i^{[\mu}\braket{\Delta\psi_i^{\nu]}}
-2i\braket{\Delta\bar\psi_i^{[\mu}}\Psi_i^{\nu]}$ 
we have
\begin{align}
\begin{aligned}
  m_iv_{i,\mu}\braket{\Delta S_i^{\mu\nu}}+
  \braket{\Delta p_{i,\mu}}S_i^{\mu\nu}&=
  -i(\braket{\Delta p_{i,\mu}}\bar\Psi_i^\mu
  +m_iv_i^\mu\braket{\Delta\bar\psi_{i,\mu}})\Psi_i^\nu\\
  &\qquad
  -i(\braket{\Delta p_{i,\mu}}\Psi_i^\mu
  +m_iv_i^\mu\braket{\Delta\psi_{i,\mu}})\bar\Psi_i^\nu\\
  &\qquad
  -im_i(v_i\cdot\bar\Psi_i\braket{\Delta\psi_i^\nu}+
  v_i\cdot\Psi_i\braket{\Delta\bar\psi_i^\nu})\,.
\end{aligned}
\end{align}
Therefore, $p_{i,\mu}S_i^{\mu\nu}$ is conserved only if,
in addition to  \eqns{eq:conB}{eq:conC},
we choose $v_i\cdot\Psi_i=v_i\cdot\bar\Psi_i=0$.
This is consistent with our observation in \sec{sec:routhians}
that the $\cN=2$ SUSY theory agrees with the spinning particle action
only when we use the covariant SSC: $\pi_{i,\mu}S_i^{\mu\nu}=0$,
which is implied by $\pi_i\cdot\psi_i=\pi_i\cdot\bar\psi_i=0$.
Meanwhile,
\begin{equation}
  {\cal S}_{i,\mu\nu}\braket{\Delta S_i^{\mu\nu}}=
  2\Psi_i\cdot\bar\Psi_i(\bar\Psi_i^\mu\braket{\Delta\psi_{i,\mu}}
  +\braket{\Delta\bar\psi_{i,\mu}}\Psi_i^\mu)\,,
\end{equation}
so preservation of the spin lengths $\tr(S_i\cdot S_i)=-2s_i^2$
is guaranteed by \eqn{eq:conD}.

One should note that the background symmetries \eqref{eq:shiftSymmetries}
are gauge fixed by our previous requirements that $b\cdot v_i=0$
and $v_i\cdot\Psi_i=v_i\cdot\bar\Psi_i=0$,
the latter implying $v_{i,\mu}\cS_i^{\mu\nu}=0$.
In terms of the shifts \eqref{eq:shiftSymmetries}
these constraints are achieved using
\begin{align}
  \begin{aligned}
    \eps_i&=v_i\cdot\Psi_i\,, \qquad
    \bar\eps_i=v_i\cdot\bar\Psi_i\,,\qquad
    \alpha_i=0\,, \\
    \xi_1&=\frac{b\cdot(\gamma v_2-v_1)-v_1\cdot\cS_2\cdot v_2+\gamma v_1\cdot\cS_1\cdot v_2}{\gamma^2-1}\,,\\
    \xi_2&=\frac{b\cdot(v_2-\gamma v_1)+v_1\cdot\cS_1\cdot v_2-\gamma v_1\cdot\cS_2\cdot v_1}{\gamma^2-1}\,.
\end{aligned}
\end{align}
However, no information is lost:
full dependence on terms of the form $b\cdot v_i$ and $v_{i,\mu}\cS_i^{\mu\nu}$
is restored to the eikonal phase by shifting
\begin{align}\label{eq:restoreTerms}
\begin{aligned}
  \cS_i^{\mu\nu}&\to\cS_i^{\mu\nu}+2(v_i\cdot\cS_i)^{[\mu}v_i^{\nu]}\,,\\
  b^\mu&\to b^\mu+\xi_2v_2^\mu-\xi_1v_1^\mu
  +\cS_2^{\mu\nu}v_{2,\nu}-\cS_1^{\mu\nu}v_{1,\nu}\,,
\end{aligned}
\end{align}
with $\xi_i$ as given above.

Given our imposition of these background constraints,
in order to truly ``check'' the background symmetries of the eikonal phase
one should calculate it without making these requirements a priori.
Up to the 2PM order described in \sec{sec:Examples} we have done so,
as a separate calculation which involved generalizing the Feynman rules in
\sec{sec:feynmanRules} to the inclusion of such terms.
Notice also that the background field symmetries \eqref{eq:shiftSymmetries}
continue to apply when $C_{E,i}\neq0$:
although local SUSY is spoiled by the presence of
additional curvature couplings in the action,
approximate SUSY persists up to spin-squared effects as discussed in \sec{sect:approxsusy}.
So we continue to expect conservation of energy, spin length, and the SSC.
In \rcite{Jakobsen:2021smu} the same approximate SUSY was also seen acting
on the leading-PM waveform $\braket{h_{\mu\nu}(k)}$.

\section{The eikonal phase and derived observables}
\label{sec:Examples}

In this section we compute the eikonal phase
$\chi=-i\log{\cal Z}_{\rm WQFT}$ up to 2PM order,
and from it derive the momentum impulse
$\braket{\Delta p_i^\mu}$,
spin kick $\braket{\Delta S_i^{\mu\nu}}$
and aligned-spin scattering angle $\theta$
using the relationships established in \sec{sec:eikonal}.

\subsection{2PM eikonal phase}
\label{sec:eikPhase2PM}

Up to 2PM order the eikonal phase is
given as the sum of four vacuum diagrams in the WQFT:
\begin{align}\label{eq:eikonal2PM}
  i \chi &=\,\,\,\,\,\,
  \begin{tikzpicture}[baseline={(current bounding box.center)}]
  \coordinate (inA) at (-1,1);
  \coordinate (outA) at (1,1);
  \coordinate (inB) at (-1,-1);
  \coordinate (outB) at (1,-1);
  \coordinate (xA) at (0,1);
  \coordinate (xB) at (0,-1);
  \draw [dotted] (inA) -- (xA);
  \draw [dotted] (inB) -- (xB);
  \draw [dotted] (xB) -- (outB);
  \draw [dotted] (xA) -- (outA);
  \draw [fill] (xA) circle (.08);
  \draw [fill] (xB) circle (.08);
  \draw [photon] (xA) -- (xB);
  \end{tikzpicture}\\[7pt]
  &+
  \begin{tikzpicture}[baseline={(current bounding box.center)}]
  \coordinate (inA) at (-1,1);
  \coordinate (outA) at (2,1);
  \coordinate (inB) at (-1,-1);
  \coordinate (outB) at (2,-1);
  \coordinate (xA) at (0,1);
  \coordinate (yA) at (1,1);
  \coordinate (xB) at (0,-1);
  \coordinate (yB) at (1,-1);
  \draw [dotted] (inA) -- (xA);
  \draw [zParticleF] (xA) -- (yA);
  \draw [dotted] (yA) -- (outA);
  \draw [dotted] (inB) -- (xB);
  \draw [dotted] (xB) -- (yB);
  \draw [dotted] (yB) -- (outB);
  \draw [fill] (xA) circle (.08);
  \draw [fill] (yA) circle (.08);
  \draw [fill] (xB) circle (.08);
  \draw [fill] (yB) circle (.08);
  \draw [photon] (xA) -- (xB);
  \draw [photon] (yA) -- (yB);
  \end{tikzpicture}\,+\,
  \begin{tikzpicture}[baseline={(current bounding box.center)}]
  \coordinate (inA) at (-1,1);
  \coordinate (outA) at (2,1);
  \coordinate (inB) at (-1,-1);
  \coordinate (outB) at (2,-1);
  \coordinate (xA) at (0,1);
  \coordinate (yA) at (1,1);
  \coordinate (xB) at (0,-1);
  \coordinate (yB) at (1,-1);
  \draw [dotted] (inA) -- (xA);
  \draw [zParticle] (xA) -- (yA);
  \draw [dotted] (yA) -- (outA);
  \draw [dotted] (inB) -- (xB);
  \draw [dotted] (xB) -- (yB);
  \draw [dotted] (yB) -- (outB);
  \draw [fill] (xA) circle (.08);
  \draw [fill] (yA) circle (.08);
  \draw [fill] (xB) circle (.08);
  \draw [fill] (yB) circle (.08);
  \draw [photon] (xA) -- (xB);
  \draw [photon] (yA) -- (yB);
  \end{tikzpicture}\,+\,
  \begin{tikzpicture}[baseline={(current bounding box.center)}]
  \coordinate (inA) at (-1,1);
  \coordinate (outA) at (2,1);
  \coordinate (inB) at (-1,-1);
  \coordinate (outB) at (2,-1);
  \coordinate (C) at (.5,0);
  \coordinate (A) at (.5,1);
  \coordinate (xB) at (-.4,-1);
  \coordinate (yB) at (1.4,-1);
  \draw [dotted] (inA) -- (A);
  \draw [dotted] (A) -- (outA);
  \draw [dotted] (inB) -- (xB);
  \draw [dotted] (xB) -- (yB);
  \draw [dotted] (yB) -- (outB);
  \draw [fill] (C) circle (.08);
  \draw [fill] (A) circle (.08);
  \draw [fill] (xB) circle (.08);
  \draw [fill] (yB) circle (.08);
  \draw [photon] (C) -- (A);
  \draw [photon] (xB) -- (C);
  \draw [photon] (yB) -- (C);
  \end{tikzpicture}\,+\,
  \begin{tikzpicture}[baseline={(current bounding box.center)}]
  \coordinate (inA) at (-1,1);
  \coordinate (outA) at (2,1);
  \coordinate (inB) at (-1,-1);
  \coordinate (outB) at (2,-1);
  \coordinate (A) at (.5,1);
  \coordinate (xB) at (-.2,-1);
  \coordinate (yB) at (1.2,-1);
  \draw [dotted] (inA) -- (A);
  \draw [dotted] (A) -- (outA);
  \draw [dotted] (inB) -- (xB);
  \draw [dotted] (xB) -- (yB);
  \draw [dotted] (yB) -- (outB);
  \draw [fill] (A) circle (.08);
  \draw [fill] (xB) circle (.08);
  \draw [fill] (yB) circle (.08);
  \draw [photon] (xB) -- (A);
  \draw [photon] (yB) -- (A);
  \end{tikzpicture}\nn\\[5pt]
  &\qquad+ \mathcal{O}(G^3)\,,\nn
\end{align}
where mirror diagrams $(1\leftrightarrow2)$ are left implicit
and we sum over both directions of the arrowed line
(representing a propagating spin mode $\psi_i^{\prime a}$).
Explicit expressions are obtained using the Feynman rules
given in \sec{sec:feynmanRules}.
For example,
the 1PM contribution only involves the graviton source vertex \eqref{eq:vertexH}
and has the explicit form
\begin{align}\label{eq:leadingDefDiagram}
  \begin{tikzpicture}[baseline={(current bounding box.center)}]
  \coordinate (inA) at (-1,1);
  \coordinate (outA) at (1,1);
  \coordinate (inB) at (-1,-1);
  \coordinate (outB) at (1,-1);
  \coordinate (xA) at (0,1);
  \coordinate (xB) at (0,-1);
  \draw [dotted] (inA) -- (xA);
  \draw [dotted] (inB) -- (xB);
  \draw [dotted] (xB) -- (outB);
  \draw [dotted] (xA) -- (outA);
  \draw [fill] (xA) circle (.08);
  \draw [fill] (xB) circle (.08);
  \draw [photon] (xA) -- (xB) node [midway, left] {$q\!\uparrow$};
  \draw (inA) node [left] {$1$};
  \draw (inB) node [left] {$2$};
  \end{tikzpicture}&=i\frac{\kappa^2m_1 m_2}{4}\!
  \int_q e^{iq\cdot b}\dd(q\cdot v_1)\dd(q\cdot v_2)
  \frac{P_{\mu\nu;\rho\sigma}}{q^2+i\eps}\\[-25pt]
  &\,\,\times\!\big(v_1^\mu v_1^\nu-i(q\cdot\cS_1)^\mu v_1^\nu
    -\sfrac12(q\cdot\cS_1)^{\mu}(q\cdot\cS_1)^{\nu}
    +\sfrac{C_{E,1}}2v_1^\mu v_1^\nu(q\cdot\cS_1\cdot\cS_1\cdot q)
  \big)\nn\\
  &\,\,\times\!\big(v_2^\rho v_2^\sigma+i(q\cdot\cS_2)^\rho v_2^\sigma
    -\sfrac12(q\cdot\cS_2)^{\rho}(q\cdot\cS_2)^{\sigma}
    +\sfrac{C_{E,2}}2v_2^\rho v_2^\sigma(q\cdot\cS_2\cdot\cS_2\cdot q)
  \big),\nn
\end{align}
where we discard all terms above $\cO(\cS^2)$; $b^\mu=b_2^\mu-b_1^\mu$
and we integrate over the off-shell momentum $q$ of the exchanged graviton.
Similar expressions are easily assembled for the other diagrams;
for a worldline propagator of either type ($z^\mu$ or $\psi'^\mu$)
we perform a one-dimensional integral $\int_\omega$ over the
intermediate energy $\omega$.

An important practical consideration is our use of the constant spinors $\Psi_i^\mu$:
in general, we prefer final results to be expressed in terms of the
physically relevant antisymmetric spin tensors
$\cS_i^{\mu\nu}=-2i\bar{\Psi}_i^{[\mu}\Psi_i^{\nu]}$.
This motivates our writing the interaction vertices
\eqref{eq:vertexH}--\eqref{eq:vertexHS}
in terms of $\cS^{\mu\nu}$ wherever possible ---
so that most of the graphs in \eqn{eq:eikonal2PM}
depend on $\Psi_i^\mu$ only via $\cS_i^{\mu\nu}$.
The only exception is the third diagram,
which carries an overall factor $\bar{\Psi}_1^\mu\Psi_1^\nu$
due to the spinor vertex \eqref{eq:vertexHS}
(appropriately contracted with the rest of the expression).
Manifest dependence on $\cS_1^{\mu\nu}$ is only recovered once the counterpart
diagram with the arrowed line pointing in the opposite direction is included:
except for its overall dependence on $\Psi_1^\mu$ the expression is identical,
and we recover $\bar{\Psi}_1^\mu\Psi_1^\nu+\Psi_1^\mu\bar{\Psi}_1^\nu=i\cS_1^{\mu\nu}$
as an overall factor.

As the techniques used to integrate these expressions are now well-established
(see e.g.~\rcites{Cristofoli:2020uzm,Kalin:2020mvi}) we relegate those details to \app{sec:integrals},
and here simply present our results.
As explained in \sec{sec:eikonal},
$\chi=\sum_{n=1}^\infty G^n\chi^{(n)}$ depends only on
the orthogonal components of $b^\mu$ and $\cS_i^{\mu\nu}$
with respect to the velocities $v_i^\mu$,
so we set $b\cdot v_i=0$ and $(v_i\cdot\cS_i)^\mu=0$ (the covariant SSC)
without loss of generality.
At 1PM order the various $D$-dimensional contributions are\footnote{The
  zeroth-order-spin contribution \eqref{eq:1PMnospin}
  is logarithmically divergent in $D=4$ dimensions ---
  this pole is unphysical and affects neither the impulse nor spin kick.
}
\begin{subequations}\label{eq:1PMresult}
  \begin{align}
    \left.\chi^{(1)}\right|_{\cS_1^0\cS_2^0}&=
    \frac{2\pi^{2-\sfrac{D}2}\Gamma(\sfrac{D}2-2)((D-2)\gamma^2-1)m_1m_2}
    {(D-2)|b|^{D-4}\sqrt{\gamma^2-1}}\,,\label{eq:1PMnospin}\\
    %%%
    \left.\chi^{(1)}\right|_{\cS_1\cS_2^0}&=
    \frac{4\pi^{2-\sfrac{D}2}\Gamma(\sfrac{D}2-1)\gamma m_1m_2}{|b|^{D-3}\sqrt{\gamma^2-1}}\,
    \bH\cdot\cS_1\cdot v_2\,,\\
    %%%
    \left.\chi^{(1)}\right|_{\cS_1\cS_2}&=
    \frac{2\pi^{2-\sfrac{D}2}\Gamma(\sfrac{D}2-1)m_1m_2}{|b|^{D-2}(\gamma^2-1)^{3/2}}\\
    &\quad\times
    \Big(
    (\gamma^2-1)
    \big(
    \gamma\tr(\cS_1\cdot\cS_2)
    -
    (D-2)
    (
    \bH\cdot\cS_1\cdot v_2\,\bH\cdot\cS_2\cdot v_1
    -
    \gamma \bH\cdot\cS_1\cdot\cS_2\cdot \bH
    )
    \big)
    \nn\\&\quad\quad
    -
    v_2\cdot\cS_1\cdot\cS_2\cdot v_1
    \Big)
    \,,\nn\\
    %%%
    \left.\chi^{(1)}\right|_{\cS_1^2\cS_2^0}&=
    \frac{2\pi^{2-\sfrac{D}2}\Gamma(\sfrac{D}2-1)m_1m_2}{(D-2)|b|^{D-2}(\gamma^2-1)^{3/2}}\\
    &\qquad\times
    \Big(
    (\gamma^2-1)
    \big(
    (D-2)^2
    (\bH\cdot\cS_1\cdot v_2)^2
    +
    (D-2)
    \bH\cdot\cS_1\cdot\cS_1\cdot\bH-2s_1^2
    \big)
    \nn\\&\qquad\qquad
    +
    \big(
    D-1-\gamma^2(D-2)
    \big)
    v_2\cdot\cS_1\cdot\cS_1\cdot v_2
    \nn\\&\qquad\qquad
    -C_{E,1}
    \big(
    (D-2)\gamma^2-1
    \big)
    \Big(
    (\gamma^2-1)
    \big(
    (D-2)\bH\cdot\cS_1\cdot\cS_1\cdot\bH-2s_1^2
    \big)
    \nn\\ &\qquad\qquad\qquad
    +v_2\cdot\cS_1\cdot\cS_1\cdot v_2
    \Big)
    \Big)
    \,,\nn
\end{align}
\end{subequations}
where $\bH^\mu:=b^{\mu}/|b|$, $|b|=\sqrt{-b\cdot b}$
and we recall that $\gamma=v_{1}\cdot v_{2}$.
As the 2PM results are more involved we provide them here only in $D=4$ dimensions:
\begin{subequations}\label{eq:2PMresult}
  \begin{align}
      \left.\chi^{(2)}\right|_{\cS_1^0\cS_2^0}&=
      \frac{3\pi(5\gamma^2-1)(m_1+m_2)m_1m_2}{4|b|\sqrt{\gamma^2-1}}\,,\\
      %%%
      \left.\chi^{(2)}\right|_{\cS_1\cS_2^0}&=
      \frac{\pi\gamma(5\gamma^2-3)(4m_1+3m_2)m_1m_2}{4|b|^2(\gamma^2-1)^{3/2}}
      \bH\cdot\cS_1\cdot v_2\,,\\
      %%%
      \left.\chi^{(2)}\right|_{\cS_1\cS_2}&=
      \frac{\pi(m_1+m_2)m_1m_2}{4|b|^3(\gamma^2-1)^{5/2}}\\
      &\qquad\times\big((\gamma^2-1)(\gamma(5\gamma^2-3)(2\tr(\cS_1\cdot\cS_2)
      +3\bH\cdot\cS_1\cdot\cS_2\cdot\bH)\nn\\
      &\qquad\qquad-
      9(5\gamma^2-1)\bH\cdot\cS_1\cdot v_2\,\bH\cdot\cS_2\cdot v_1)
      -3(3\gamma^2-1)v_2\cdot\cS_1\cdot\cS_2\cdot v_1\big)\,,\nn\\
      %%%
      \left.\chi^{(2)}\right|_{\cS_1^2\cS_2^0}&=
      \frac{\pi\,m_1m_2}{64|b|^3(\gamma^2-1)^{5/2}}\\
      \times\big(\,\,\,\,\,\,\,\,\,\,\,
      &\!\!\!\!\!\!\!\!\!\!\!
      8(\gamma^2-1)((13\gamma^4-42\gamma^2+21)m_1-4(3\gamma^2-1)m_2)s_1^2\nn\\
      &\!\!\!\!\!\!\!\!\!\!\!
      -6(\gamma^2-1)((29\gamma^4-66\gamma^2+29)m_1-4(3\gamma^2-1)m_2)
      \bH\cdot\cS_1\cdot\cS_1\cdot\bH\nn\\
      &\!\!\!\!\!\!\!\!\!\!\!
      +24(\gamma^2-1)((31\gamma^2-11)m_1+3(5\gamma^2-1)m_2)
      (\bH\cdot\cS_1\cdot v_2)^2\nn\\
      &\!\!\!\!\!\!\!\!\!\!\!
      -6((49\gamma^4-90\gamma^2+33)m_1+4(5\gamma^4-9\gamma^2+2)m_2)
      v_2\cdot\cS_1\cdot\cS_1\cdot v_2\nn\\
      &\!\!\!\!\!\!\!\!\!\!\!
      +4C_{E,1}(\gamma^2-1)
      ((125\gamma^4-138\gamma^2+29)m_1+2(45\gamma^4-42\gamma^2+5)m_2)s_1^2\nn\\
      &\!\!\!\!\!\!\!\!\!\!\!
      -3C_{E,1}(\gamma^2-1)
      ((155\gamma^4-174\gamma^2+35)m_1+4(30\gamma^4-29\gamma^2+3)m_2)
      \bH\cdot\cS_1\cdot\cS_1\cdot\bH\nn\\
      &\!\!\!\!\!\!\!\!\!\!\!
      -3C_{E,1}((95\gamma^4-102\gamma^2+23)m_1+4(15\gamma^4-13\gamma^2+2)m_2))
      v_2\cdot\cS_1\cdot\cS_1\cdot v_2
      \big)\,.\nn
  \end{align}
\end{subequations}
We have confirmed agreement between these four-dimensional results and
\rcites{Bern:2020buy,Kosmopoulos:2021zoq}.\footnote{
	We find it simplest to compare our expressions with covariant scattering
	amplitudes before the Fourier transform with respect to $q^\mu$ is taken.
}
For the full $D$-dimensional expressions at 2PM order we refer the reader to
the ancillary file attached to the \texttt{arXiv} submission of this paper.

\subsection{1PM observables}
\label{sec:1PMobservables}

The impulse, spin kick and aligned-spin scattering angle
are derived from the eikonal phase $\chi$ using
\begin{align}\label{eq:derivatives}
	\braket{\Delta p_{i,\mu}}=
  -\frac{\partial\chi}{\partial b_i^\mu}\,, &&
  \braket{\Delta S_i^{\mu\nu}}
  =\frac4{m_i}{\cal S}_i^{\rho[\mu}
  \frac{\partial\chi}{\partial{{\cal S}_{i,\nu]}}^\rho}\,, &&
  \sin\!\left(\frac\theta2\right)=
  -\frac1{2p_\infty}\frac{\partial\chi}{\partial|b|}\,,
\end{align}
where $p_\infty$ is the centre-of-mass momentum \eqref{eq:angleCoM}.
Care should be taken with these derivatives as the 1PM eikonal phase
given in \eqn{eq:1PMresult}
depends only on the $(D-2)$-dimensional
orthogonal components of $b^\mu=b_2^\mu-b_1^\mu$ and $\cS_i^{\mu\nu}$
with respect to the velocities $v_i^\mu$.
As explained in \sec{sec:eikonal},
full dependence on terms of the form $b\cdot v_i$ and $(v_i\cdot\cS_i)^\mu$
is restored to the eikonal phase using the SUSY shifts in \eqn{eq:restoreTerms},
after which the derivatives \eqref{eq:derivatives} may be taken without issue.
One may then safely re-impose $b\cdot v_i=0$ and $(v_i\cdot\cS_i)^\mu=0$
on the resulting physical observables.
Notice that \eqn{eq:derivatives} implies conservation of momentum
$\braket{\Delta p_2^\mu}=
-\braket{\Delta p_1^\mu}$ as all dependence on $b_i^\mu$
comes via the relative impact parameter $b^\mu=b_2^\mu-b_1^\mu$
(we are free to choose a spacetime origin).
At higher-PM orders this implies that the scattering is \emph{conservative},
i.e.~by this procedure we miss radiation-reaction effects for which
$\braket{\Delta p_2^\mu}\neq
-\braket{\Delta p_1^\mu}$.

The $\cO(G)$ part of the impulse
$\braket{\Delta p_1^\mu}=\sum_{n=1}^\infty G^n\braket{\Delta p_1^\mu}^{(n)}$
is given by
\begin{subequations}
  \begin{align}
    &\left.\braket{\Delta p_1^\mu}^{(1)}\right|_{\cS_1^0\cS_2^0}=
    \frac{
      4\pi^{2-\sfrac{D}2}\Gamma(\sfrac{D}2-1)((D-2)\gamma^2-1)m_1 m_2
    }{
      (D-2)|b|^{D-3}\sqrt{\gamma^2-1}
    }\hat{b}^\mu\,,\\
    %%%%%%DELTAP_S1
    &\left.\braket{\Delta p_1^\mu}^{(1)}\right|_{\cS_1\cS_2^0}=
    -\frac{
      4\pi^{2-\sfrac{D}2}\Gamma(\sfrac{D}2-1)\gamma m_1m_2
    }{
      |b|^{D-2}\sqrt{\gamma^2-1}
    }\,
    \left((D-2)\bH\cdot\cS_1\cdot v_2\,\hat{b}^\mu+(\cS_1\cdot v_2)^\mu\right)\,,\\
    %%%%%%DELTAP_S1S2
    &\left.\braket{\Delta p_1^\mu}^{(1)}\right|_{\cS_1\cS_2}=
    \frac{
      4\pi^{2-\sfrac{D}2}\Gamma(\sfrac{D}2)m_1m_2
    }{
      |b|^{D-1}\sqrt{\gamma^2-1}
    }
    \bigg(
    \bH\cdot\cS_1\cdot v_2 (v_1\cdot\cS_2)^\mu
    +
    \bH\cdot\cS_2\cdot v_1 (v_2\cdot\cS_1)^\mu
    \nn
    \\
    &\qquad
    +\Big(
    \gamma D \bH\cdot\cS_1\cdot\cS_2\cdot\bH
    -D\bH\cdot\cS_1\cdot v_2 \bH\cdot\cS_2\cdot v_1
    +\gamma\tr(\cS_1\cdot\cS_2)
    -\frac{v_2\cdot\cS_1\cdot\cS_2\cdot v_1}{\gamma^2-1}
    \Big)\bH^\mu
    \nn
    \\
    &\qquad
    +\gamma(\bH\cdot\cS_1\cdot\cS_2\cdot P_{12})^\mu
    +\gamma(P_{12}\cdot\cS_1\cdot\cS_2\cdot \bH)^\mu
    \bigg)
    \ ,
    \\
    %%%%%%DELTAP_S1^2
    &\left.\braket{\Delta p_1^\mu}^{(1)}\right|_{\cS_1^2\cS_2^0}=
    \frac{
      2\pi^{2-\sfrac{D}2}\Gamma(\sfrac{D}2-1)m_1m_2
    }{
      |b|^{D-1}\sqrt{\gamma^2-1}
    }
    \bigg(
    -2(D-2)\bH\cdot\cS_1\cdot v_2 (v_2\cdot \cS_1)^\mu
    \nn
    \\
    &\qquad
    +
    \Big(
    (D-2)D(\bH\cdot\cS_1\cdot v_2)^2
    +
    D\bH\cdot\cS_1\cdot\cS_1\cdot\bH
    +
    \frac{D-1-(D-2)\gamma^2}{\gamma^2-1}v_2\cdot\cS_1\cdot\cS_1\cdot v_2
    \nn
    \\
    &\qquad
        -2s_1^2
    \Big)
    \bH^\mu
    +2 (\bH\cdot\cS_1\cdot\cS_1\cdot P_{12})^\mu
    -
    C_{E,1}\big((D-2)\gamma^2-1\big)
    \Big(
    2(\bH\cdot\cS_1\cdot\cS_1\cdot P_{12})^\mu
    \nn
    \\
    &\qquad
    +
    \big(
    D\bH\cdot\cS_1\cdot\cS_1\cdot\bH
    +
    \tr(\cS_1\cdot P_{12}\cdot\cS_1)
    \big)\bH^\mu
    \Big)
    \bigg)
    \ ,
\end{align}
\end{subequations}
and $\braket{\Delta p_2^\mu}=-\braket{\Delta p_1^\mu}$.
These expressions are manifestly orthogonal to $v_i^\mu$,
as required by \eqn{eq:conA},
which is apparent given our use of the projector $P^{\mu\nu}_{12}$
to the $(D-2)$-dimensional space orthogonal to these velocities:
\begin{align}\label{eq:proj12}
  P_{12}^{\mu\nu} &=
  \eta^{\mu\nu} + \frac1{\gamma^2-1} 
  \left[ v_1^\mu v_1^\nu - 2 \gamma v_1^{(\mu} v_2^{\nu)} + v_2^\mu v_2^\nu \right]\,.
\end{align}
The $\cO(G)$ part of the spin kick
$\braket{\Delta S_1^{\mu\nu}}=
\sum_{n=1}^\infty G^n\braket{\Delta S_1^{\mu\nu}}^{(n)}$ is
\begin{subequations}
  \begin{align}
    %%%%%%DELTA S _S1
    &\left.\braket{\Delta S_1^{\mu\nu}}^{(1)}\right|_{\cS_1^1\cS_2^0}=
    -\frac{
      8\pi^{2-\sfrac{D}2}\Gamma(\sfrac{D}2-1) m_2
    }{
      |b|^{D-3}\sqrt{\gamma^2-1}
    }
    \\
    &\qquad\times
    \bigg(
    \frac{(\bH\cdot\cS_1)^{[\mu}{v_1}^{\nu]}}{D-2}
    +\gamma(v_2\cdot\cS_1)^{[\mu}{\bH}^{\nu]}
    -\gamma(\bH\cdot\cS_1)^{[\mu}{v_2}^{\nu]}
    \bigg)
    \ ,
    \nn
    \\
    %%%%%%%DELTA S _S1^2
    &\left.\braket{\Delta S_1^{\mu\nu}}^{(1)}\right|_{\cS_1^2\cS_2^0}=
    \frac{
      8\pi^{2-\sfrac{D}2}\Gamma(\sfrac{D}2-1) m_2
    }{
      |b|^{D-2}\sqrt{\gamma^2-1}
    }
    \\
    &\qquad\times
    \bigg(
    -(\cS_1\cdot\cS_1\cdot\bH)^{[\mu} {\bH}^{\nu]}
    -(D-2)\bH\cdot\cS_1\cdot v_2
    \Big(
    (\cS_1\cdot v_2)^{[\mu}{\bH}^{\nu]}
    -(\cS_1\cdot\bH)^{[\mu} {v_2}^{\nu]}
    \Big)
    \nn
    \\
    &\qquad\qquad
    +\frac{\gamma (\cS_1\cdot\cS_1\cdot v_2)^{[\mu} {v_1}^{\nu]}
    }{(\gamma^2-1)(D-2)}
    +\frac{1+(D-2)\gamma^2-D}{(\gamma^2-1)(D-2)}
    (\cS_1\cdot\cS_1\cdot v_2)^{[\mu} {v_2}^{\nu]}
    \nn
    \\
    &\qquad\qquad
    +
    C_{E,1}\big((D-2)\gamma^2-1\big)
    \Big(
    (\cS_1\cdot\cS_1\cdot\bH)^{[\mu}{\bH}^{\nu]}
    -
    \frac{(\cS_1\cdot\cS_1\cdot v_2)^{[\mu}(\gamma v_1 - v_2)^{\nu]}}
    {(\gamma^2-1)(D-2)}
    \Big)
    \bigg)
    \ ,
    \nn
    \\
    %%%%%%%DELTA S _S1S2
    &\left.\braket{\Delta S_1^{\mu\nu}}^{(1)}\right|_{\cS_1^1\cS_2^1}=
    \frac{
      4\pi^{2-\sfrac{D}2}\Gamma(\sfrac{D}2-1) m_2
    }{
      |b|^{D-2}\sqrt{\gamma^2-1}
    }
    \\
    &\qquad\times
    \bigg(
    \gamma(D-2)
    \big(
    (\cS_1\cdot\bH)^{[\mu} (\cS_2\cdot\bH)^{\nu]}
    -(\cS_1\cdot\cS_2\cdot\bH)^{[\mu}{\bH}^{\nu]}
    \big)
    -\frac{(\cS_1\cdot v_2)^{[\mu} (\cS_2\cdot v_1)^{\nu]}}{\gamma^2-1}
    \nn
    \\
    &\qquad\qquad
    -\frac{(\cS_1\cdot\cS_2\cdot v_1)^{[\mu}
        (\gamma v_1 - v_2)^{\nu]}}{\gamma^2-1}
    -2\gamma (\cS_1\cdot\cS_2)^{[\mu\nu]}
    \nn
    \\
    &\qquad\qquad
    -(D-2)
    \big(
    (\cS_1\cdot\bH)^{[\mu}{v_2}^{\nu]} 
    -
    (\cS_1\cdot v_2)^{[\mu}{\bH}^{\nu]} 
    \big)
    \bH\cdot\cS_2\cdot v_1
    \bigg)
    \ .
    \nn
  \end{align}
\end{subequations}
The other components are $\braket{\Delta\cS_1^{\mu\nu}}^{(1)}|_{\cS_1^0\cS_2^1}=
\braket{\Delta\cS_1^{\mu\nu}}^{(1)}|_{\cS_1^0\cS_2^2}=0$;
$\braket{\Delta\cS_2^{\mu\nu}}^{(1)}$ is recovered by simple relabelling.
Finally, the $\cO(G)$ part of the scattering angle
$\theta=\sum_{n=1}^\infty G^n\theta^{(n)}$ emerging in the case of aligned spins is
\begin{align}
\begin{aligned}
  &
  \theta^{(1)}
  =
  \frac{2\pi^{2-\frac{D}2}(D-3) \Gamma(\frac{D}{2}-1)E
  }{
     |b|^{D-3}(\gamma^2-1)
  }
  \bigg(
  \frac{
    2(D-2)\gamma^2-2
  }{
    (D-3)(D-2)
  }
  +
  2\gamma\sqrt{\gamma^2-1}
  \frac{s_1+s_2}{|b|}
  \\
  &\qquad\qquad
  +
  \big(
  (D-2)\gamma^2 -\sfrac34 D +2
  \big)
  \frac{(s_1+s_2)^2}{|b|^2}
  -
  \frac{D-4}{4}\frac{(s_1-s_2)^2}{|b|^2}
  \\
  &\qquad\qquad
  -
  \big(
  (D-2)\gamma^2-1
  \big)
  \left(
    \frac{C_{E,1}s_1^2+C_{E,2}s_2^2}{|b|^2}
  \right)
  \bigg)\,.
\end{aligned}
\end{align}
To specify aligned spins we have inserted
\begin{align}\label{eq:alignedSpin}
  \cS_1^{\mu\nu}=
  2
  s_1\frac{b^{[\mu} (\gamma v_1-v_2)^{\nu]}}{|b|\sqrt{\gamma^2-1}}\,,&&
  \cS_2^{\mu\nu}=
  2
  s_2\frac{b^{[\mu} (v_1-\gamma v_2)^{\nu]}}{|b|\sqrt{\gamma^2-1}}\,,
\end{align}
with the normalizations ensuring that
$\tr(\cS_i\cdot\cS_i)=-2s_i^2$.
The aligned-spin tensors live in the subspace spanned by $b^\mu$, $v_1^\mu$ and $v_2^\mu$, which together with the SSC $(v_i\cdot\cS_i)^\mu=0$ defines them uniquely.
This definition ensures planar dynamics and includes the conventional definition in four spacetime dimensions.

\subsection{2PM observables}
\label{sec:2PMobservables}

The 2PM momentum impulse and spin kick are again derived from
the eikonal phase by taking derivatives with respect to the
background parameters $b_i^\mu$ and $\cS_i^{\mu\nu}$ \eqref{eq:derivatives};
however, an additional subtlety is the interpretation of these background
parameters (and $v_i^\mu$).
In general, we prefer to express observables in terms of
$b_{\pm\infty}^\mu$, $v_{i,\pm\infty}^\mu$ and $\cS_{i,\pm\infty}^\mu$
taken in the far past or future:
\begin{subequations}
\begin{align}
  x_i^\mu(\tau)&\xrightarrow[\tau\to\pm\infty]{}b^\mu_{i,\pm\infty}+\tau v^\mu_{i,\pm\infty}\,,\\
  S_i^{\mu\nu}(\tau)&\xrightarrow[\tau\to\pm\infty]{}\cS^{\mu\nu}_{i,\pm\infty}\,.
\end{align}
\end{subequations}
With the time-symmetric worldline propagators \eqref{eq:wlPropagators}
the currently used background parameters are
$b^\mu=\sfrac12(b_{+\infty}^\mu+b_{-\infty}^\mu)+\cO(G^2)$,
$v_i^\mu=\sfrac12(v_{i,+\infty}^\mu+v_{i,-\infty}^\mu)+\cO(G^2)$
and $\cS_i^{\mu\nu}=\sfrac12(\cS_{i,+\infty}^{\mu\nu}+\cS_{i,-\infty}^{\mu\nu})+\cO(G^2)$.
To leading order in $G$ the transition is straightforwardly
accomplished using $\braket{\Delta b^\mu}$,
the momentum impulse
$\braket{\Delta p_i^\mu}=m_i(v_{i,+\infty}^\mu-v_{i,-\infty}^\mu)$ and the 
spin kick $\braket{\Delta S_i^{\mu\nu}}=
\cS_{i,+\infty}^{\mu\nu}-\cS_{i,-\infty}^{\mu\nu}$:
\begin{subequations}\label{eq:eikParams}
\begin{align}
b^\mu_{\pm\infty}&=b^\mu\pm\frac{\braket{\Delta b^\mu}}{2}+\cO(G^2)\,,\\
v^\mu_{i,\pm\infty}&=v^\mu_i\pm\frac{\braket{\Delta p_i^\mu}}{2m_i}+\cO(G^2)\,,\\
\cS^{\mu\nu}_{i,\pm\infty}&=\cS^{\mu\nu}_i\pm\frac{\braket{\Delta\cS_i^{\mu\nu}}}2+\cO(G^2)\,.
\end{align}
\end{subequations}
Using $0=v_i^\mu\braket{\Delta p_{i,\mu}}$ \eqref{eq:conA}
it also follows that  $\gamma_{\pm\infty}=\gamma+\cO(G^2)$.
Given our 1PM results in \sec{sec:1PMobservables}
we lack only the 1PM expression for $\braket{\Delta b^\mu}$;
those results are unaffected as the parameters differ by terms $\cO(G)$.

Conservation of angular momentum at 1PM order
$\braket{\Delta J^{\mu\nu}}=
{\cal J}^{\mu\nu}_{+\infty}-{\cal J}^{\mu\nu}_{-\infty}=0$
gives us an $\cO(G)$ expression for $\braket{\Delta b^\mu}$.
The total angular momentum at future/past infinity is
\begin{align}
\begin{aligned}
  {\cal J}_{\pm\infty}^{\mu\nu}&=\sum_{i=1}^2
  m_i\Big(2 b_{i,\pm\infty}^{[\mu} v_{i,\pm\infty}^{\nu]}+\cS_{i,\pm\infty}^{\mu\nu}\Big)\\
  &=
  2(b_{1,\pm\infty}+b_{2,\pm\infty})^{[\mu}P^{\nu]}+2q_{\pm\infty}^{[\mu}b_{\pm\infty}^{\nu]}+
  \sum_{i=1}^2m_i{\cal S}_{i,\pm\infty}^{\mu\nu}\,,
\end{aligned}
\end{align}
where $P^\mu:=\frac12(m_1v^\mu_{1,\pm\infty}+m_2v^\mu_{2,\pm\infty})$
and $q_{\pm\infty}^\mu:=\frac12(m_1v^\mu_{1,\pm\infty}-m_2v^\mu_{2,\pm\infty})$.
We note that
$\braket{\Delta p_1^\mu}=
-\braket{\Delta p_2^\mu}=q^\mu_{+\infty}-q^\mu_{-\infty}$
due to conservation of linear momentum.
To isolate the term depending on the relative impact parameter $b^\mu=b_2^\mu-b_1^\mu$
we introduce a projector to the $(D-1)$-dimensional space
orthogonal to $P^\mu$:
\begin{align}
  {\Lambda^\mu}_\nu:={\delta^\mu}_\nu-\frac{P^\mu P_\nu}{P^2}\,.
\end{align}
In effect, contraction with this projector specializes us to the
spacelike components of ${\cal J}_{\pm\infty}^{\mu\nu}$ in the center-of-mass frame:
\begin{align}
  (\Lambda \cdot{\cal J}_{\pm\infty}\cdot \Lambda)^{\mu\nu}
  =
  2
  (\Lambda\cdot q_{\pm\infty})^{[\mu} (\Lambda \cdot b_{\pm\infty})^{\nu]}
  + 
  \sum_{i=1}^2
  m_i (\Lambda\cdot{\cal S}_{i,\pm\infty}\cdot \Lambda)^{\mu\nu}\,.
\end{align}
Contracting ${\Lambda^\mu}_\rho{\Lambda^\nu}_\sigma\braket{\Delta J^{\rho\sigma}}=0$
with $q^\mu=\sfrac12(m_1v_1^\mu-m_2v_2^\mu)$ 
we rearrange to find an $\cO(G)$ expression for $\braket{\Delta b^\mu}$:
\begin{align}
  \begin{aligned}
    \braket{\Delta b^\mu}^{(1)}
    =
    \frac{\Lambda^{\mu\nu}}{q\cdot \Lambda\cdot q}
    \left(
    -
    q_\nu
    b^{\rho}
    \braket{\Delta p_{1,\rho}}^{(1)}+
    \sum_{i=1}^2
    m_i 
    \braket{\Delta S_{i,\nu\rho}}^{(1)}
    \Lambda^{\rho\sigma} q_\sigma
    \right)\,,
  \end{aligned}
  \end{align}
having inserted the expressions for $b^\mu_{\pm\infty}$,
$v^\mu_{i,\pm\infty}$ and ${\cal S}^{\mu\nu}_{i,\pm\infty}$ \eqref{eq:eikParams}.
We have also used the requirement that $b_{\pm\infty}\cdot v_{i,\pm\infty}=0$,
which using \eqn{eq:eikParams} implies $P_\mu\braket{\Delta b^\mu}=0$
and $q_\mu\braket{\Delta b^\mu}=-b_\mu\braket{\Delta p_1^\mu}$.

Our preference is to re-express observables in terms of background parameters
taken in the far past $\tau\to-\infty$;
the switch should be performed only \emph{after}
taking derivatives of the eikonal \eqref{eq:derivatives}.
The 2PM observables then pick up corrections from the 1PM observables
in \sec{sec:1PMobservables}.
As the results are quite lengthy we provide them here only in $D=4$ dimensions
up to linear order in spin;
for the full $D$-dimensional quadratic-in-spin expressions
we refer the reader to the accompanying ancillary file.
Firstly, the momentum impulse:
\begin{subequations}
\begin{align}
  \left.\braket{\Delta p^{\mu}_1}^{(2)}\right|_{\cS_1^0\cS_2^0}
  &=
  \frac{m_1m_2}{|b|^2}\bigg(
  \frac{3\pi(5\gamma^2-1)(m_1+m_2)}{4\sqrt{\gamma^2-1}}\hat{b}^\mu\\
  &\qquad\qquad\qquad
  -2\frac{(2\gamma^2-1)^2}{(\gamma^2-1)^2}
  \big(
  (\gamma m_1+m_2)v_1^\mu-(\gamma m_2+m_1)v_2^\mu
  \big)
  \bigg)
  \,,
  \nn
  \\
  \left.\braket{\Delta p^{\mu}_1}^{(2)}\right|_{\cS_1^1\cS_2^0}&=
  \frac{m_1 m_2}{|b|^2}
  \bigg(
    \frac{\hat b \cdot \cS_1 \cdot v_2}{|b|}
    \Big(
  -
  \frac{3\pi\gamma(5\gamma^2-3)}{4(\gamma^2-1)^{3/2}}
  (4m_1+3m_2)
  \hat b^\mu
  \\
  &\!\!\!\!\!\!\!\!\!\!\!\!\!\!\!\!\!\!\!\!\!\!\!\!\!\!\!\!\!\!
  +
  \frac{16\gamma^2(2\gamma^2-1)m_1+2\gamma(12\gamma^2-5)m_2}{(\gamma^2-1)^2}
  v_1^\mu
  %% \\
  %% &\!\!\!\!\!\!\!\!\!\!\!\!\!\!\!\!\!\!
  -
  \frac{16\gamma(2\gamma^2-1)m_1+2(8\gamma^4-1)m_2}{(\gamma^2-1)^2}
  v_2^\mu
  \Big)
  \nn
  \\
  &\!\!\!\!\!\!\!\!\!\!\!\!\!\!\!\!\!\!\!\!\!\!\!\!\!\!\!\!\!\!
  -\frac{2(4\gamma^2-1)m_2+8\gamma(2\gamma^2-1)m_1}{\gamma^2-1}
  \frac{(\hat b\cdot \cS_1)^\mu}{|b|}
  %% \\
  %% &\!\!\!\!\!\!\!\!\!\!\!\!\!\!\!\!\!\!
  +\frac{\pi\gamma(5\gamma^2-3)(4m_1+3m_2)}{4(\gamma^2-1)^{3/2}}
  \frac{(v_2\cdot\cS_1)^\mu}{|b|}
  \bigg)\,.
  \nn
\end{align}
\end{subequations}
Here, and from this point on,
the $-\infty$ subscripts on $\hat{b}_{-\infty}^\mu$, $v_{i,-\infty}^\mu$
and $\cS^{\mu\nu}_{i,-\infty}$ should be considered implicit.
The $\cO(G^2)$ part of the spin kick is
\begin{align}
  &
  \braket{\Delta S^{\mu\nu}_1}^{(2)}
  =
  \frac{m_2^2}{|b|^2 (\gamma^2-1)}
  \bigg(
  4\antic{(\bH\cdot\cS_1)}{{\bH}}
  -
  16 \gamma \bH\cdot\cS_1\cdot v_2
  (
  2\gamma v_2-v_1
  )^{[\mu}{\bH}^{\nu]}
  \\
  &\qquad
  -
  16\gamma^2
  \antic{(v_2\cdot\cS_1)}{{v_2}}
  +
  \frac{
    \pi\gamma(5\gamma^2-3)(4m_1+3m_2)
  }{
    2\sqrt{\gamma^2-1} m_2
  }
  \Big(
  \antic{(\bH\cdot\cS_1)}{{v_2}}
  -\antic{(v_2\cdot\cS_1)}{{\bH}}
  \Big)
  \nn
  \\
  &\qquad
  -
  \frac{
    \pi(5\gamma^4+6\gamma^2-3)m_1
    +
    3\pi(3\gamma^2-1)m_2
  }{
    2\sqrt{\gamma^2-1}m_2
  }
  \antic{(\bH\cdot\cS_1)}{{v_1}}
  \nn
  \\
  &\qquad
  +
  \frac{
    4(2\gamma^2-1)^2m_1
    +
    4\gamma(4\gamma^2-3)m_2
  }{
    (\gamma^2-1)m_2
  }
  \antic{(v_2\cdot\cS_1)}{{v_1}}
  \bigg)
  \nn
\end{align}
When expressed in terms of the new background parameters,
different relationships are satisfied by these observables:
\begin{subequations}
  \begin{align}
    m_i^2v_{i}^2&=(m_iv^\mu_{i}+\braket{\Delta p^\mu_i})^2\,,\\
    (\cS^{\mu\nu}_{i})^2&=(\cS^{\mu\nu}_{i}+\braket{\Delta S^{\mu\nu}_i})^2\,,\\
    m_iv_{i,\mu}\cS_{i}^{\mu\nu}&=
    (m_iv_{i,\mu}+\braket{\Delta p_{i,\mu}})
    (\cS^{\mu\nu}_{i}+\braket{\Delta S^{\mu\nu}_i})\,.
  \end{align}
\end{subequations}
However, the interpretation is still the same:
supercharges are conserved between initial and final states.
Lastly, the 2PM scattering angle in $D=4$ is
\begin{align}
    &
    \theta^{(2)}
    =\frac{E(m_1+m_2)}{|b|^2}\bigg(
    \frac{
      3\pi (5\gamma^2-1)
    }{
      4(\gamma^2-1)
    }
    +
    \frac{\pi\gamma(5\gamma^2-3)
    }{
      2(\gamma^2-1)^{3/2}
    }
    \Big(
    \frac{3m_2+4m_1}{m_1+m_2}
    \frac{s_1}{|b|}+(1\leftrightarrow2)
    \Big)
    \nn
    \\
    &\ \ 
    +
    \frac{
      3\pi
    }{
      2(\gamma^2-1)^2|b|^2
    }
    \bigg[
      \frac{
      (95\gamma^4-102\gamma^2+15)m_1
      +
      4\big(
      15\gamma^4-15\gamma^2+2
      \big)
      m_2
      }{8(m_1+m_2)}
      s_1^2
      +(1\leftrightarrow2)
      \nn
      \\
      &\qquad\qquad\quad
      -C_{E,1}
      \frac{
      (125\gamma^4-138\gamma^2+29)m_1
      +
      2(45\gamma^4-42\gamma^2+5)m_2
      }{16(m_1+m_2)}
      s_1^2
      +(1\leftrightarrow2)
      \nn
      \\
      &\qquad\qquad\quad
      +(20\gamma^4-21\gamma^2+3)
      s_1s_2
      \bigg]\bigg)\,,
\end{align}
where the specialization to aligned spins was given earlier \eqref{eq:alignedSpin}.
The spin-free part of our 2PM scattering angle in $D$ dimensions
(provided in the ancillary file) agrees with earlier results
\cite{Cristofoli:2020uzm,KoemansCollado:2019ggb}.

Finally, we have observed that the 2PM eikonal phase $\chi$ presented in
\sec{sec:eikPhase2PM} is invariant under the transformations
between intermediate and past background parameters \eqref{eq:eikParams} ---
unlike the momentum impulse $\braket{\Delta p_1^\mu}$
and spin kick $\braket{\Delta S_1^{\mu\nu}}$ presented above.
One may therefore freely replace $b^\mu\to b_{\pm\infty}^\mu$,
$v_i^\mu\to v^\mu_{i,\pm\infty}$ and
${\cal S}^{\mu\nu}_i\to{\cal S}^{\mu\nu}_{i,\pm\infty}$
in the expressions \eqref{eq:1PMresult} and \eqref{eq:2PMresult}
without changing their validity.
However, one should then exercise caution when deriving physical observables:
the simple derivatives of the eikonal phase \eqref{eq:derivatives}
apply only to the ``averaged'' background parameters,
rather than those defined at past/future infinity \eqref{eq:eikParams}.
Instead, following the approach conjectured in \rcites{Bern:2020buy,Kosmopoulos:2021zoq},
it is appropriate to include terms quadratic in the eikonal phase.
Noting that the shifts \eqref{eq:eikParams} cannot be considered
instances of the background field symmetries discussed in \sec{sect:sym}
(e.g.~the background symmetries \eqref{eq:shiftSymmetries} leave $v_i^\mu$ invariant)
the additional invariance here suggests the existence of a larger class of conserved
quantities than those limited here to the individual worldlines.
We leave this tantalizing question for future work.

\section{Conclusions}

The $\cN=2$ supersymmetric worldline action provides an alternative
description of a compact spinning object
up to terms quadratic in spin $\cO(\cS^2)$ (quadrupoles).
Using this equivalence we have  shown how quadratic-in-spin effects
may be incorporated into the worldline quantum field theory (WQFT)
prescription for scattering massive bodies in a curved background \cite{Mogull:2020sak}.
The classical spin tensors $S_i^{ab}=-2i\bar\psi_i^{[a}\psi_i^{b]}$
(in a local frame $e_a^\mu$) are considered composite fields,
built from the complex Grassmann-valued vectors $\psi_i^a$
living on each worldline $i$.
Conveniently, this provides for a Lagrangian worldline formalism
that involves neither a body-fixed frame nor angular velocity tensor.
The technology was previously used to obtain
the far-field time-domain waveform from a scattering of two massive bodies
(black holes, neutron stars or stars) to leading order in $G$ \cite{Jakobsen:2021lvp};
here we elaborated on it further.

While Kerr black holes are privileged,
and represented by the unique $\cN=2$ supersymmetric theory,
finite-size effects may also be incorporated starting at $\cO(\cS^2)$
by adding terms that only preserve SUSY approximately (up to $\cO(\psi^{5})$).
The conserved supercharges have natural physical interpretations:
conservation of energy, spin length
and the spin-supplementary condition (SSC) $p_\mu S^{\mu\nu}=0$
along each worldline.
While these are conserved locally in the supersymmetric theory,
with the inclusion of finite-size effects they are only conserved  approximately up to $\mathcal{O}(\cS^3)$.
The analogue can be seen
 in \rcite{Jakobsen:2021lvp},
where the time-domain waveform is approximately supersymmetric when finite-size
effects are included and exactly supersymmetric in the Kerr-black hole case.

Our main result is an explicit expression for the $D$-dimensional eikonal phase
$\chi=-i\log{{\cal Z}_{\rm WQFT}}$ up to $\cO(G^2)$ (2PM order),
where ${\cal Z}_{\rm WQFT}$ is the partition function of the WQFT.
This was obtained as a sum of tree-level vacuum diagrams
integrated over the momenta (or energies on the worldlines) of internal lines.
From the eikonal phase we showed how one may derive three key observables:
the momentum impulse $\braket{\Delta p_1^\mu}=-\braket{\Delta p_2^\mu}$,
spin kicks $\braket{\Delta S_i^{\mu\nu}}$
and (for aligned spins) the scattering angle $\theta$.
In $D=4$ dimensions these observables agree with previous results
\cite{Kosmopoulos:2021zoq,Liu:2021zxr}.
The requirements of energy conservation and
preservation of both the spin-supplementary condition (SSC)
and spin length follow naturally from the supersymmetry.
Another important subtlety is the interpretation of background parameters
$b_i^\mu$, $v_i^\mu$ and $\cS_i^{\mu\nu}$:
in the eikonal phase these are defined at an intermediate point of the scattering,
and an interpolation is needed to relate them to those in the far past ($\tau\to-\infty$).

Our work offers numerous follow-up opportunities,
which we will explore enthusiastically.
Naturally, one wonders about the prospects for extending the formalism
beyond quadratic order in the spins.
As explained in \sec{sec:spinOnWorldline}, in a flat spacetime background
there exist $\cN$-supersymmetric worldline theories
with real Grassmann-valued vectors $\psi_\alpha^a(\tau)$
carrying flavor indices $\alpha=1,\ldots,\cN$
that generically describe the propagation of spin-$\cN/2$ particles
\cite{Brink:1976sz,Howe:1988ft,Howe:1989vn}.
The main obstacle is generalizing these theories to an
arbitrary curved spacetime background whilst preserving supersymmetry.
Yet we have seen that perturbative deformations of the supercharges yielding
an approximate supersymmetry are possible; it would be worthwhile to revisit
the issue under these premises.
Fortunately, the higher spin limitation does not exist in gauge theories:
the so-called $\sqrt{\rm Kerr}$ theory \cite{Arkani-Hamed:2019ymq},
which enjoys a complex worldsheet description \cite{Guevara:2020xjx},
is a natural candidate for study.
Given ongoing research on the double copy in WQFT \cite{Shi:2021qsb},
this could provide a window on higher spins for the Kerr black hole and is left for future work.

We also see excellent prospects for applying the spinning
WQFT formalism to higher-PM order calculations.
As explained in \rcite{Mogull:2020sak} we are not limited to using the eikonal phase:
we can also compute $\braket{\Delta p_i^\mu}$ directly by drawing
graphs with an outgoing deflection mode $z_i^\mu$;
similarly we can obtain $\braket{\Delta\psi_i^\mu}$,
and therefore $\braket{\Delta S_i^{\mu\nu}}$ via \eqn{eq:spinKickderiv}, 
by drawing graphs with an outgoing $\psi_i^{\prime\mu}$ line.
There already has been excellent progress at 3PM order in the non-spinning case
\cite{Kalin:2020fhe,DiVecchia:2021bdo,Brandhuber:2021eyq,Herrmann:2021tct} 
including radiation reaction effects \cite{Damour:2020tta,Bini:2021gat}.

Finally, in the non-spinning case a link between scalar-graviton S-matrix elements
and operator expectation values in the WQFT has been formally provided
by a worldline path integral ``Feynman-Schwinger'' representation
of the graviton-dressed scalar propagator \cite{Mogull:2020sak}.
We would like to extend this link to include spin effects ---
again, gauge theory theory will provide a useful starting point
given that the $n$-dressed electron propagator is already known
\cite{Ahmadiniaz:2020wlm,Ahmadiniaz:2021gsd}
(see \rcites{Schubert:2001he,Edwards:2019eby,Edwards:2021elz} for comprehensive reviews).
In gravity, such a dressed propagator is not currently known,
and when obtained will provide for a complete theoretical map between the
different PM-based approaches to spinning black hole scattering.

%%%%%%%%%%%%%%%%%%%%%%%%%%%%%%%%%%%%%%%%%%%%%%%%%%%%
\begin{acknowledgments}
  We would like to thank Roberto Bonezzi, Alessandra Buonanno, Paolo Pichini, Christian Schubert and Justin Vines for helpful discussions.
  We are also grateful for use of Gregor K\"alin's C\texttt{++} graph library.
  GUJ's and GM's research is funded by the Deutsche Forschungsgemeinschaft (DFG, German Research Foundation),
  Projektnummer 417533893/GRK2575 ``Rethinking Quantum Field Theory''.
\end{acknowledgments}
%%%%%%%%%%%%%%%%%%%%%%%%%%%%%%%%%%%%%%%%%%%%%%%%%%%%

\appendix
\section{Supersymmetry}\label{AppA}

The relevant part of the $e=1/m$
gauge-fixed $\mathcal{N}=2$ worldline action in curved space \eqref{eq:finalAction}
reads
\begin{align}\label{eq:finalActionGauged}
S= -m\int\!\d\tau \Bigl [&\sfrac{1}{2}g_{\mu\nu}\dot x^{\mu}\dot x^{\nu}
+ i\bar\psi^{a}\dot\psi_{a}+i\dot x^{\mu}\omega_{\mu ab}\bar\psi^{a}\psi^{b}  + \sfrac{1}{2} R_{abcd}\bar\psi^{a}\psi^{b}\bar\psi^{c}\psi^{d}
\Bigr ]\, .
\end{align}
and we now want to prove its SUSY invariance.
The SUSY transformations of $x^{\mu}$, $\psi^{a}$ and $\bar\psi^{a}$ were quoted in \eqn{N=2SUSYcurvedfinal}:
\begin{align}\label{eq:susyvarapp}
\delta x^{\mu}&= ie^{\mu}_{a} (\bar\epsilon\psi^{a} + \epsilon\bar\psi^{a})  \, , 
\quad 
\delta \psi^{a} = -\epsilon e^{a}_{\mu} \, \dot x^{\mu} - \delta x^{\mu}\omega_{\mu}{}^{a}{}_{b}\psi^{b}\, ,
\end{align}
and are augmented by
\be
\delta e^{a}_{\mu}= \partial_{\nu}e^{a}_{\mu}\, \delta x^{\nu}\, , \quad
\delta \omega_{\mu ab}= \partial_{\nu}  \omega_{\mu ab}\, \delta x^{\nu}\, , \quad
\delta R_{abcd}= \partial_{\nu}R_{abcd}\,\delta x^{\nu}\, .
\ee 
In order to show the invariance of the action $S$  we analyze the variation $\delta S$
order-by-order in the fermions $\psi^{a}$. At linear order only
the variations of the first three terms in \eqn{eq:finalActionGauged} contribute and one
 finds
\begin{align}
\delta S|_{\text{lin}}&= -m\int d\tau \Bigl (
g_{\mu\nu}\dot x^{\mu}i\bar\epsilon \dot\psi^{\nu}
 -\dot x ^{\mu}e^{a}_{\mu}\, i \bar\epsilon \dot\psi^{a}
 % \nn\\ & +
+ \left [ \sfrac{1}{2} e^{\rho}_{a}\partial_{\rho}g_{\mu\nu} -\omega_{(\mu \nu) a }\right ] i\bar\epsilon \psi^{a}
\dot x^{\mu}\dot x^{\nu} + \text{c.c}\Bigr )\, .
\end{align}
Using $\sfrac{1}{2} e^{\rho}_{a}\partial_{\rho}g_{\mu\nu} -\omega_{(\mu \nu) a }=
- g_{\rho(\mu} \partial_{\nu) }e_{a}^{\rho}$
the last term is rewritten as $- g_{\rho\mu}\dot x^{\mu}(\frac{d}{d \tau} 
e^{\rho}_{a}) i\bar\epsilon \psi^{a}$. Noting that the first two terms combine
to the same expression, but an opposite sign, we see the vanishing of the linear in
$\psi^{a}$ variation $\delta S|_{\text{lin}}$. At cubic order one picks up contributions
only from the last three terms in \eqn{eq:finalActionGauged} and finds
\begin{align}
\delta S|_{\text{cubic}}&= -m\int d\tau \Bigl (
[\partial_{\rho} \omega_{\mu ab} - \partial_{\mu}\omega_{\rho a b}-\omega_{\mu ac}\, \omega_{\rho}{}^{c}{}_{b}+\omega_{\rho ac}\, \omega_{\mu}{}^{c}{}_{b} + R_{\mu\rho ab}]
i\dot x^{\mu}\delta x^{\rho}\bar\psi^{a}\psi^{b}\, ,
\end{align}
which vanishes identically using the spin-connection based definition of the Riemann-tensor. At quintic order one merely considers the variation of the last term
in \eqn{eq:finalActionGauged}. Here the three types of contributions conspire to yield
\begin{align}
\delta & S|_{\text{quintic}}=\\
-&m\int d\tau  [ \partial_{\mu}R_{abcd}+ \omega_{\mu a }{}^{e} R_{ebcd}
+ \omega_{\mu b }{}^{e} R_{aecd}+ \omega_{\mu c }{}^{e} R_{abed}
+ \omega_{\mu d }{}^{e} R_{abce}]\, \delta x^{\mu} \bar\psi^{a}\psi^{b}\bar\psi^{c}\psi^{d}
\, , \nn
\end{align}
which constitutes the covariant derivative of the Riemann-tensor $\nabla_{\mu} R_{abcd}$. By virtue of $\nabla_{\mu}e^{a}_{\nu}=0$ we need to show the vanishing of the term
$\nabla_{\mu} R_{\alpha\nu\beta\rho} \, \psi^{\mu}\psi^{\nu}\psi^{\rho}$ (and analogously for 
$\psi\to \bar\psi$). Using the cyclicity of the Riemann tensor
in its last three indices, $R_{\alpha\nu\beta\rho}=-R_{\alpha\beta\rho\nu}
-R_{\alpha\rho\nu\beta}$,
 and the anti-commuativity of the $\psi$'s we have
\be
\nabla_{\mu} R_{\alpha\nu\beta\rho} \, \psi^{\mu}\psi^{\nu}\psi^{\rho}
= \frac{1}{2} \nabla_{\mu} R_{\alpha\beta\nu\rho} \, \psi^{\mu}\psi^{\nu}\psi^{\rho}\, .
\ee
This expression vanishes by virtue of the Biancchi identity
$\nabla_{\mu} R_{\alpha\beta\nu\rho}+\nabla_{\nu} R_{\alpha\beta\rho\mu}
+\nabla_{\rho} R_{\alpha\beta\mu\nu}=0$.

For completeness, let us now also look at the supersymmetry variation of the finite-size term \eqref{eq:finiteSizeDef} relevant for (neutron) stars:
\begin{equation}\label{eq:SEapp}
  S_{E}=
  -m\,C_{E}\int\!\d\tau\,  R_{a\mu b\nu}\dot{x}^\mu\dot{x}^\nu \bar\psi^a \psi^b\, 
  \, P_{cd}\, \bar\psi^{c}\,\psi^{d}\,.
\end{equation}
Varying this under \eqref{eq:susyvarapp} we produce terms of order three and
five in the worldline fermions $\psi^{a}$. The order-five terms would also receive
contributions from putative order-six terms (yielding spin${}^{3}$ effects) in the effective worldline theory that we
have not considered. Hence, of relevance here are only the order-three terms in the supersymmetry variation of \eqref{eq:SEapp}, which in fact come close to vanishing:
varying the first two fermions in $S_{E}$ yields zero due to
$R_{\rho\mu b\nu}\dot x^{\mu}\dot x^{\nu}\dot x^{\rho}=0$. One is left with
\be
\delta S_{E}\Bigr|_{\psi^{3}}=
  m\,C_{E}\int\!\d\tau\,  R_{a\mu b\nu}\dot{x}^\mu\dot{x}^\nu \bar\psi^a \psi^b\, 
 g_{\rho\sigma} (
 \bar\epsilon \, P_{\rho\sigma}\psi^{\rho} \dot x^{\sigma}
 + \epsilon \, P_{\rho\sigma}\bar\psi^{\rho} \dot x^{\sigma}
 )\,.
\ee
The terms in the bracket vanish by virtue of the projector property: $P_{\rho \sigma}\xdot^{\sigma}=0$. Hence the finite-size term $S_{E}$ is supersymmetric approximately, i.e.
\be
\delta S_{E}= \cO(\psi^{5}) \, .
\ee
As we would need to include a new layer of $\psi^{6}$ terms in order to describe
spinning massive objects at the spin-cubed order, the SUSY variation of these not-considered
terms would induce $\cO(\psi^{5})$ terms which would talk to the above.

\section{Integrals}\label{sec:integrals}

To integrate the 1PM contribution to the eikonal phase $\chi$
given in \eqn{eq:leadingDefDiagram} we require
expressions for the following class of $D$-dimensional Fourier transforms:
\begin{align}
  I^{\mu_1\mu_2\ldots\mu_n}_\nu(D):=
  \int_qe^{iq\cdot b}\,\dd(q\cdot v_1)\dd(q\cdot v_2)|q|^\nu
  q^{\mu_1}q^{\mu_2}\cdots q^{\mu_n}\,,
\end{align}
where $|q|^2=-q\cdot q$ and $n\leq2$;\footnote{An
  $\cO(\cS^\alpha)$ contribution generically requires integrals with rank $n=\alpha$.
}
the $\nu\neq-1$ generalization becomes relevant at 2PM order.
The scalar integral is straightforwardly evaluated in a $(D-2)$-dimensional
space orthogonal to $v_1^\mu$ and $v_2^\mu$
(see e.g.~\rcite{Mogull:2020sak}) with the well-known result:
\begin{align}
  I_\nu(D)=
  \frac{2^\nu}{\pi^{(D-2)/2}\sqrt{\gamma^2-1}}
  \frac{\Gamma(\sfrac{D-2+\nu}2)}{\Gamma(-\sfrac{\nu}2)}
  \left(-b\cdot P_{12}\cdot b\right)^{-\frac{D-2+\nu}2}\,,
\end{align}
where the projector $P^{\mu\nu}_{12}$ to the $(D-2)$-dimensional space
orthogonal to $v_i^\mu$ was given in \eqn{eq:proj12}.
The generalization to higher-rank integrals follows easily
by taking derivatives with respect to $b^\mu$:
\begin{equation}
  I^{\mu_1\mu_2\ldots\mu_n}_\nu(D)=
  (-i)^n\frac{\partial^n}
  {\partial b_{(\mu_1}\partial b_{\mu_2}\cdots\partial b_{\mu_n)}}I_\nu(D)\,.
\end{equation}
It is important not to impose $b\cdot v_i=0$
until after these derivatives have been taken ---
hence our use of the projector $P_{12}^{\mu\nu}$.

To integrate the various contributions appearing at 2PM order we
additionally require full knowledge of the following family of integrals:
\begin{align}
  J_{\nu_1,\nu_2,\nu_3}^{\mu_1\mu_2\ldots\mu_n}(D):=
  \int_\ell\frac{\dd(\ell\cdot v_1)}
  {(\ell^2+i\eps)^{\nu_1}((\ell-q)^2+i\eps)^{\nu_2}(\ell\cdot v_2+i\eps)^{\nu_3}}
  \ell^{\mu_1}\ell^{\mu_2}\cdots\ell^{\mu_n}\,,
\end{align}
with $v_1^\mu\leftrightarrow v_2^\mu$ related by symmetry
and $n\leq3$.
The scalar integral is straightforwardly evaluated by choosing
the rest frame of the first body, $v_1^\mu=(1,\mathbf{0})$,
and performing the resulting $(D-1)$-dimensional one-loop integral:
\begin{equation}
  J_{\nu_1,\nu_2,\nu_3}(D)=
%  i^{\nu_3}
  %  (-1)^{\nu_1+\nu_2+\nu_3}
  (-i)^{2\nu_1+2\nu_2+\nu_3}
  (4\pi)^{\frac{1-D}2}
  \left(\frac{4}{\gamma^2-1}\right)^{\sfrac{\nu_3}2}
  \Gamma_{\nu_1,\nu_2,\nu_3}(D-1)\,
  |q|^{D-1-2\nu_1-2\nu_2-\nu_3}
  \,,
\end{equation}
where
\begin{equation}
  \Gamma_{\nu_1,\nu_2,\nu_3}(D):=
  \frac{\Gamma(\nu_1+\nu_2+\sfrac{\nu_3}2-\sfrac{D}2)\Gamma(\sfrac{\nu_3}2)}
  {2\Gamma(\nu_1)\Gamma(\nu_2)\Gamma(\nu_3)}
  \frac{\Gamma(\sfrac{D}2-\nu_1-\sfrac{\nu_3}2)\Gamma(\sfrac{D}2-\nu_2-\sfrac{\nu_3}2)}
  {\Gamma(D-\nu_1-\nu_2-\nu_3)}\,.
\end{equation}
When $\nu_3=0$ we use $\Gamma(\sfrac{\nu_3}2)=2\Gamma(\nu_3)$.\footnote{As
  internal self-consistency checks we also found
  it helpful to consider integration-by-parts (IBP) relations between these integrals
  using \texttt{LiteRed} \cite{Lee:2012cn,Lee:2013mka}
  and \texttt{FIRE6} \cite{Smirnov:2019qkx}.
}
Higher-rank integrals are expanded on a basis of tensors
living in the $(D-1)$-dimensional space orthogonal to $v_1^\mu$:
namely $P_1^{\mu\nu}$, $P_1^{\mu\nu}v_{2,\nu}$, and $q^\mu$,
where $P_1^{\mu\nu}:=\eta^{\mu\nu}-v_1^\mu v_1^\nu$
is the projector orthogonal to $v_1^\mu$.
The coefficients are found by contracting an ansatz with these tensors,
and for example with $n=1$ one can easily show that
\begin{equation}
  J^\mu_{\nu_1,\nu_2,\nu_3}=\frac{q^\mu}{2|q|^2}
  \left(|q|^2J_{\nu_1,\nu_2,\nu_3}-J_{\nu_1-1,\nu_2,\nu_3}+J_{\nu_1,\nu_2-1,\nu_3}\right)+
  \frac{P_1^{\mu\nu}v_{2,\nu}}{\gamma^2-1}J_{\nu_1,\nu_2,\nu_3-1}\,,
\end{equation}
which holds in any dimension $D$.
Similar relations hold for $n=2$ and $n=3$.

\bibliographystyle{JHEP}
\bibliography{../bib/wqft_spin}

\end{document}